\documentclass[]{interact}

\usepackage{amsmath}
\usepackage{amssymb}
\usepackage{amsfonts}
\usepackage{empheq}

\usepackage[linesnumbered,ruled,vlined]{algorithm2e}
\let\oldnl\nl
\newcommand{\nonl}{\renewcommand{\nl}{\let\nl\oldnl}}

\usepackage{enumitem} 

\usepackage{epsfig}

\usepackage{csvsimple}

\usepackage{hyperref}

\usepackage[usenames]{color}

\usepackage{longtable}

\usepackage{pdflscape}

\usepackage{colortbl}
\definecolor{tableShade}{rgb}{0.945,0.961,0.980}

\usepackage{fancyvrb}

\usepackage{shorttoc}

\usepackage[titles]{tocloft}

\usepackage{multicol}

\usepackage{cleveref}

\makeatletter
\renewcommand*\cleardoublepage{\clearpage\if@twoside
\ifodd\c@page\else
\hbox{}
\thispagestyle{empty}
\newpage
\if@twocolumn\hbox{}\newpage\fi\fi\fi}
\makeatother

\newcommand{\sfrac}[2]{\mathchoice
  {\kern0em\raise.5ex\hbox{\the\scriptfont0 #1}\kern-.15em/
   \kern-.15em\lower.25ex\hbox{\the\scriptfont0 #2}}
  {\kern0em\raise.5ex\hbox{\the\scriptfont0 #1}\kern-.15em/
   \kern-.15em\lower.25ex\hbox{\the\scriptfont0 #2}}
  {\kern0em\raise.5ex\hbox{\the\scriptscriptfont0 #1}\kern-.2em/
   \kern-.15em\lower.25ex\hbox{\the\scriptscriptfont0 #2}}
  {#1\!/#2}}

\usepackage{listings}

\definecolor{gray}{rgb}       {0.8,0.8,0.8}
\definecolor{light-blue}{rgb} {0.8,0.8,1.0}
\definecolor{light-green}{rgb}{0.8,1.0,0.8}
\definecolor{light-red}{rgb}  {1.0,0.9,0.9}

\newcommand{\Dmix}{{\cal D}_{m}}
\newcommand{\Vdiff}{{\cal V}}

\newcommand{\EF}{\textit{ef}}
\newcommand{\Wbar}{\overline{W}}

\newcommand{\Uvec}{\hspace{-.1em}\textbf{{\em U}}}

\newcommand{\Nav}{N_{\textbf{\hspace{-.2em}\scalebox{.75}{{\em A}}}}}

\newcommand{\dvec}{\textbf{{\em d}} \hspace{.05em}}

\newcommand{\fluxm}{\mathbf{\Gamma}_m}
\newcommand{\fluxe}{\mathbf{\Gamma}_e}

\begin{document} 


\baselineskip16pt


\title{A spectral deferred correction strategy for low Mach number reacting flows subject to electric fields}

\author{
\name{L. Esclapez\thanks{Contact: L. Esclapez. E-mail:  lesclapez@lbl.gov}, V. Ricchiuti, J. B. Bell and M. S. Day}
\affil{Center for Computational Sciences and Engineering, Lawrence Berkeley National Laboratory, Berkeley, CA 94720, USA}
}

\maketitle

\begin{abstract}

\noindent
We propose an algorithm for low Mach number reacting flows subjected to electric field that includes the chemical production and transport of charged species. This work is an extension of a multi-implicit spectral deferred correction (MISDC) algorithm  designed to advance the conservation equations in time at scales associated with advective transport. The fast and nontrivial interactions of electrons with the electric field are treated implicitly using a Jacobian-Free Newton Krylov approach for which a preconditioning strategy is developed. Within the MISDC framework, this enables a close and stable coupling of diffusion, reactions and dielectric relaxation terms with advective transport and is shown to exhibit second-order convergence in space and time. The algorithm is then applied to a series of steady and unsteady problems to demonstrate its capability and stability. Although developed in a one-dimensional case, the algorithmic ingredients are carefully designed to be amenable to multidimensional applications.
 \end{abstract}


 \section{Introduction}
 
Experiments have shown that applying electric fields to flames can provide an effective control of the combustion process by enhancing flame propagation speed, improving flame stabilization and reducing pollutant emissions \cite{Fialkov:1997, Starikovskiy:2013}. However, the development of such technology has proven difficult without a clear understanding of the interaction mechanisms between the flame and the electric field, and the use of electric fields is currently limited to flame detection sensors \cite{Adams:1995}.
 
The chemical decomposition of hydrocarbons proceeds mainly through reactions involving neutral intermediate radicals. However, some reactions, called chemi-ionization reactions, also produce small quantities of charged chemical species and electrons \cite{Calcote:1957,Goodings:1979a,Goodings:1979b}. These particles undergo a force when subjected to an electric field and their interactions with the surrounding gas can result in a global flame response to the electric field. Three major effects have been advanced in the literature \cite{Fialkov:1997}: 1) the collision of charged particles with neutral ones induces a bulk convective transport in the gas called the ionic wind effect; 2) the transport of highly reactive charged particles from the reactive layer of the flame to the low temperature zone enhances the fuel oxidation rate; and 3) for strong electric fields ohmic heating increases the flame temperature, resulting in a higher flame speed. These processes were found to have an effect on flame speed \cite{Jaggers:1971, Tewari:1975, Marcum:2005, Won:2008, Kim:2011}, flame stabilization \cite{Cessou:2012} and NO$_x$ and soot formation \cite{Kono:1989, Vatazhin:1995,Saito:1999}. The extent to which each process is important depends on the applied potential difference, the polarity, the distance between the electrodes and the flame, and the operating conditions, making it difficult to compare results from different experiments and to provide clear design guidelines for engineers. 

Over the last decade, several groups have developed numerical methods to analyze the interactions of an electric field with charged particles in a flame. In most applications, the flame can be considered weakly ionized, i.e., the number density of electrons is much smaller than that of neutrals. However, the presence of charged particles, especially light electrons, results in challenging numerical issues associated with the wide scale separation between the electron dielectric relaxation scale and the comparatively slow hydrodynamic scale. Consequently, early studies focused mainly on steady-state one-dimensional flames \cite{Jones:1972, Pedersen:1993, Prager:2007} without an external applied electric field and identified the main chemical pathways associated with ions as well as the role of the ambipolar diffusion in the charged species spatial distribution. More recently, these steady-state numerical studies have been used to provide a more complete characterization of the flame response to external forcing (also called the $i-V$ curve, relating the current drawn from the flame to the applied voltage difference) \cite{Peerlings:2013, Speelman:2015, Belhi:2017, Han:2017}. In agreement with experimental evidence, the effect of the external electric field is found to strongly depend on its polarity. The current is found to increase linearly with the potential difference before reaching a saturation current for high (positive) voltage.
These studies highlight the dependence of the numerical results on the choices of the chemical mechanism and, to a lesser extent, on the modeling of electron and ion transport properties \cite{Bisetti:2012, Han:2015}. Steady-state multi-dimensional simulations have also been reported \cite{Papac:2008, Renzo:2018}, showing that the simulations are able to capture qualitatively the change in flame shape and position resulting from the ionic wind. Due to the aforementioned multi-scale nature of the problem, fewer unsteady simulations are reported in the literature \cite{Yamashita:2009, Belhi:2010, Belhi:2013, Belhi:2017, Belhi:2019}. These simulations capture the effect of the electric field on the flame base position and investigate both direct current (DC) and alternative current (AC) conditions. To partially alleviate the fast electron drift velocity constraint on the stability of the numerical method, Belhi \emph{et al.} \cite{Belhi:2010, Belhi:2013} employed a small value of the electron mobility $\kappa_e$ and a linearized approximation of the charged species transport equation. The effects of these assumptions on the flame response was not evaluated and this approach cannot be extended to more realistic values of $\kappa_e$ and higher intensity external electric fields without significant reduction of the simulation time step. In the plasma community, semi-implicit methods have been developed to overcome the electron time scale constraint \cite{Hagelaar:2000}. However these approaches allow at best a couple orders of magnitude increase of the time step ($\sim 10^{-11}-10^{-13}$ s depending on the intensity of the electric field), which remains several order of magnitudes smaller than the hydrodynamic time scale in typical turbulent combustion applications ($\sim 10^{-7}-10^{-8}$ s).

In this paper, we propose a strategy based on multi-implicit spectral deferred correction (MISDC) method \cite{Nonaka:2012} to include the coupling between charged species and an electric field in a low Mach number combustion framework. The MISDC approach allows tight coupling between the different physical processes in a multi-scale simulation by including the effect of each process in their separate integration (in contrast to Strang splitting methods that consider each process sequentially and independently \cite{Nonaka:2012}). To alleviate the electron dielectric relaxation time scale constraint, the non-linear system formed by the coupled electron conservation equation and electrostatic potential equation is solved implicitly using a preconditioned Jacobian-free Newton Krylov (JFNK) method.
 
The paper is organized as follows. In Section 2 we introduce the low Mach number conservation equations including the electrostatic potential equation as well as the chemical and transport models. In Section 3 we discuss the changes implemented in the MISDC algorithm and details of the solution of the implicit non-linear system. We then provide a skeletal description of the time advance procedure. In Section 4 we present results for premixed flames in 1D under DC and AC conditions. Finally, the paper finishes with the main take-away of our approach and discusses future work. 

\section{Low Mach number equation set}

\subsection{Low Mach number equation set}

This paper builds on the low Mach number equations set reported in previous work \cite{Day:2000, Nonaka:2012}, with the addition of an electrical drift contribution in the momentum, species and enthalpy equations, a separate conservation equation for the electron number density and a Poisson equation for the electrostatic potential to obtain an electric field consistent with the charged species distribution.

In the low Mach number regime, the characteristic velocity of the fluid $\Uvec_{adv}$ is much smaller than the speed of sound $\mathbf{a}$ (typically $|\Uvec_{adv}|/a= M \sim 0.1$ or even smaller), so the effect of acoustic wave propagation can be neglected since it does not affect the dynamics of the system. In numerical simulations, this effect is mathematically removed from the equations of motion and the system evolves subject to a time step based on the advective CFL condition. In low Mach number conditions, the total pressure can be decomposed into a spatially uniform (thermodynamic) component $p_0$, and a perturbational term, $\pi$, that drives the flow:
\begin{equation}
p(\mathbf{x},t) = p_0 + \pi(\mathbf{x},t)
\end{equation}
Although the formulation supports a time varying $p_0$ (arising, for example, in closed chamber applications \cite{Nonaka:2018}), we assume an open domain here to simplify the exposition.

The set of equations describing species, electrons, enthalpy and momentum conservation in the low Mach number limit \cite{Day:2000} are given by: 
\begin{align}
\frac{\partial (\rho Y_m)}{\partial t} + \boldsymbol{\nabla} \cdot {(\Uvec_{adv} \rho Y_m) }  & = - \boldsymbol{\nabla} \cdot {\fluxm}
 +\dot{\omega}_m \qquad m=1:N \label{eq:species} \\
 \frac{\partial (n_e)}{\partial t} + \boldsymbol{\nabla} \cdot {(\Uvec_{adv} n_e) } & =  - \boldsymbol{\nabla} \cdot {\fluxe}
 +\dot{\omega}_e  \label{eq:elec} \\
 \frac{\partial (\rho h)}{\partial t} + \boldsymbol{\nabla} \cdot {(\Uvec_{adv} \rho h) } & = \boldsymbol{\nabla} \cdot {\lambda \boldsymbol{\nabla T} }
 - \sum_{m}{\boldsymbol{\nabla}  \cdot ({h_m \fluxm}}) + \sum_{m+e} z_m Y_m \fluxm \cdot \boldsymbol{E} \label{eq:nrg}\\
\frac{\partial ( \rho\Uvec_{adv})}{\partial t} + \boldsymbol{\nabla} \cdot (\rho \Uvec_{adv} \Uvec_{adv}) & = - \boldsymbol{\nabla \pi} + \boldsymbol{\nabla}  \cdot \boldsymbol{\tau}
 + \rho \sum_{m+e} z_m Y_{m} \boldsymbol{E} \label{eq:mom}
\end{align}
where $N$ is the total number of species (excluding the electrons), $\rho$ is the density, $Y_m$ is the mass fraction of species $m$, $n_e$ is the electron number density, $\Uvec_{adv}$ is the fluid advective velocity, $\fluxm$ (resp. $\fluxe$) is the diffusion mass flux of species $m$ (electrons), $h= \sum_m (Y_m h_m)$ is the mixture total (sensible and chemical) enthalpy with $h_m(T)$ the enthalpy of species $m$, $\dot{\omega}_m$ (resp. $\dot{\omega}_e$) is the production rate of species $m$ (electrons) due to chemical reactions, $\lambda$ is the thermal conductivity, $z_m$ is the electric charge per unit mass of species $m$, $\boldsymbol{E}$ is the electric field, and $\pi$ is the perturbational pressure arising from the low Mach number approximation. The evolution equations are closed by an equation of state, $p_0 = p(\rho,T,Y)$ for the thermodynamic pressure. Note that the low Mach assumption requires that the flow evolve subject to a constant $p_0$. This DAE system can be solved by differentiating the equation of state in the frame of the fluid and requiring that the evolution be constrained to satisfy constant pressure in this frame \cite{Majda:1985}. Here we assume a mix of ideal gases:
\begin{equation}
p_0 = \rho \frac{R_u}{\Wbar} T = \rho R_u \sum_m\frac{Y_m}{W_m}\label{eq:EOS}
\end {equation}
where $p_0$ is the ambient pressure, $\Wbar$ is the mean molecular weight of the mixture, $W_m$ is the molecular weight of species $m$ and $R_u$ is the universal gas constant. Expanding in partial derivatives and using the conservation equations, the constant $p_0$ condition can be recast as a constraint on the velocity \cite{Day:2000}:
\begin{alignat}{1}
\begin{split}
\boldsymbol{\nabla} \cdot \boldsymbol{U}_{adv}  & = \frac{1}{\rho c_p T} \bigg( \boldsymbol{\nabla} \cdot \lambda \boldsymbol{\nabla T} + \sum_m \fluxm \cdot \boldsymbol{\nabla h_m} + \sum_{m+e} z_m Y_m \fluxm \cdot \boldsymbol{E}\bigg) \\
 & + \frac{1}{\rho} \sum_m \frac{\Wbar}{W_m} \boldsymbol{\nabla} \cdot \fluxm + \frac{1}{\rho} \sum_m \bigg( \frac{\Wbar}{W_m} - \frac{h_m}{c_p T} \bigg) \dot{\omega}_m \equiv S
 \end{split}\label{eq:constr1}
\end{alignat}
where $c_p$ is the specific heat at constant pressure for the mixture. Since this constraint is a linearization of the equation of state, the thermodynamic variables will not remain consistent with $p_0$ numerically; in order to prevent this thermodynamic drift, a correction term $\delta \chi$ has been added to to the constraint equation \ref{eq:constr1}:
\begin{equation}
\boldsymbol{\nabla} \cdot \boldsymbol{U}_{adv} = S + \underbrace{\frac{f}{p_{eos}} \bigg( \frac{p_{eos} - p_0}{\Delta t}+ \boldsymbol{U}_{adv} \cdot \boldsymbol{\nabla p_{eos}} \bigg)}_{\delta \chi} \equiv \hat{S}\label{eq:divunew}
\end{equation}
where $0 < f < 1$ is a damping factor (see Day \emph{et al.} \cite{Day:2000} for details on the iterative implementation of this equation). Compared to classical low Mach number reactive flows, two additional source terms appear in the conservation equations:  1) the Lorentz volumetric forces (last term in Eq.~(\ref{eq:mom}) and 2) the ohmic heating, corresponding to the work of the Lorentz forces (last term in Eq.~(\ref{eq:nrg})).

The stress tensor in the momentum Eq.~\ref{eq:mom} is defined as:
\begin{equation}
\boldsymbol{\tau} = \mu \bigg[ \boldsymbol{\nabla} \Uvec_{adv} + (\boldsymbol{\nabla} \Uvec_{adv})^T - \frac{2}{3} \mathcal{I} (\boldsymbol{\nabla} \cdot \Uvec_{adv}) \bigg]
\end{equation}
where $\mu(Y_m,T)$ is the dynamic viscosity and $\mathcal{I}$ is the identity tensor (we ignore the bulk viscosity here).
Since neither species diffusion nor chemistry redistribute total mass, we have $\sum_m{\fluxm} = 0$ and $\sum_m{\dot{\omega}_m} = 0$. Noting that $\sum_m Y_m = 1$ (ignoring the mass of electrons), the continuity equation can be derived summing up the species continuity equations:
\begin{equation}
\frac{\partial \rho}{\partial t} + \boldsymbol{\nabla} \cdot (\rho \Uvec_{adv}) = 0
\end{equation}

The diffusion flux of species $m$ can be expressed as:
\begin{equation}
  \fluxm = \rho Y_m \boldsymbol{\Vdiff}_{m} \,, \hspace{2em}
 \boldsymbol{ \Vdiff}_{m} = - \varUpsilon_m \, \dvec_{m} = - \varUpsilon_m \, \big( \dvec_{m,d} + \dvec_{m,\EF} \big)
  = \boldsymbol{\Vdiff}_{m,d} + \boldsymbol{\Vdiff}_{m,\EF}
\end{equation}
where $\boldsymbol{\Vdiff}_{m}$ defines the diffusion velocity of species $m$ in terms of EGLIB's ``flux diffusion vector''
$\varUpsilon_m = \frac{W_m}{\Wbar} \Dmix$ \cite{ref19} and the driving forces $\dvec_m$; $\Dmix$ is the mixture-averaged diffusion coefficient of species $m$. Ignoring Dufour, Soret and barodiffusion terms, the diffusion and electric driving forces are, respectively:
\begin{equation}
	\left\{
	\begin{aligned}
		\dvec_{m,d}  & = \boldsymbol{\nabla{X}}_m \\
		\dvec_{m,\EF}  & = \frac{\nu_m q_e \Nav}{R_u T} \frac{\Wbar}{W_m} Y_m \, \boldsymbol{\nabla \phi} = z_m Y_m \frac{\Wbar}{R_u T} \boldsymbol{\nabla \phi}
	\end{aligned}\label{eq:drivingforces}
		\right.
\end{equation}
where $X_m$ is the mole fraction of species $m$, $\phi$ is the electric potential, $\nu_m$ is the valence ($e$ charges per molecule) of species $m$, $q_e$ is the elementary electron charge, $\Nav$ is Avogadro's number,  and $z_m = \nu_m q_e \Nav / W_m$ is the charge per unit mass of species $m$. 

Under the electrostatic assumption, the local electric field $\boldsymbol{E}$ is obtained from Gauss' law:
\begin{equation}
\boldsymbol{\nabla}\cdot\boldsymbol{E} = \frac{q_t}{\epsilon_0 \epsilon_r}
\label{eq:gauss}
\end{equation}
where $q_t = \sum_m z_m \rho Y_m + q_e n_e$ is the local total charge number density of the mixture and $\epsilon_0$ and $\epsilon_r=1$ are the vacuum permittivity and the relative permittivity of the gaseous medium, respectively. The electric field is the negative gradient of the electrostatic potential $\phi$, i.e:
\begin{equation}
\boldsymbol{E} = - \boldsymbol{\nabla \phi}
\label{eq:efield}
\end{equation}
Inserting Eq.~(\ref{eq:gauss}) in Eq.~(\ref{eq:efield}) we obtain the electrostatic potential equation:
\begin{equation}
 - \epsilon_0 \epsilon_r \boldsymbol{\nabla}^2 \phi = q_t \label{eq:poisson}
\end{equation}
The drift velocity $\boldsymbol{\Vdiff}_{m,ef}$ can also be written as $\boldsymbol{\Vdiff}_{m,ef} = \kappa_m \boldsymbol{E}$, where $\kappa_m$ is the mobility of species $m$ in the mixture. Thus, consistent with the Einstein relation \cite{ref20}, the mobility is defined as:
\begin{equation}
\kappa_m = \Dmix \frac{\nu_m q_e \Nav}{R_u T}
\end{equation}

The right-hand side of the diffusive driving force in Eq.~\ref{eq:drivingforces} can be rewritten as:
\begin{equation}
\boldsymbol{\nabla X}_m = \frac{\Wbar}{W_m} \boldsymbol{\nabla Y}_m + \frac{Y_m}{W_m} \boldsymbol{\nabla \Wbar}
\end{equation}
and so the diffusion fluxes can be rewritten in terms of mass fractions gradients plus $\Wbar$ corrections:
\begin{equation}
\rho Y_m \boldsymbol{\Vdiff}_{m,d} = - \rho \frac{W_m}{\Wbar} \Dmix \Big( \frac{\Wbar}{W_m} \boldsymbol{\nabla Y}_m + \frac{Y_m}{W_m} \boldsymbol{\nabla{\Wbar}} \Big) = - \rho \Dmix \boldsymbol{\nabla Y}_m - \rho \Dmix \frac{Y_m}{\Wbar} \boldsymbol{\nabla{\Wbar}}
\end{equation}
We will use this form of the transport equation to build an iterative time-implicit update scheme based on lagging the corrections and sweeping through the species with decoupled linear solves for the Crank-Nicolson update. The resulting form of the diffusive species flux will be:
\begin{equation}
\tilde{\boldsymbol{\Gamma}}_m \equiv - \rho \Dmix \boldsymbol{\nabla Y}_m - \rho \Dmix \frac{Y_m}{\Wbar} \boldsymbol{\nabla{\Wbar}} -
  \rho Y_m \kappa_m \boldsymbol{\nabla{\phi}} \label{eq:diffusionfluxes}
\end{equation}
However, since we employed mixture averaged diffusion coefficients, equation \ref{eq:diffusionfluxes} will not in general satisfy that $\sum_m \tilde{\boldsymbol{\Gamma}}_m = 0$; to conserve mass, we introduce a correction velocity \cite{Day:2000} that guarantees that these fluxes sum to zero.  Since we use an implicit approach to compute the diffusion fluxes, we first solve the implicit system to evaluate the original fluxes $\tilde{\boldsymbol{\Gamma}}_m$, then we conservatively correct $\tilde{\boldsymbol{\Gamma}}_m$ so that they sum to zero on each cell face (we will denote the corrected fluxes as $\fluxm$), and finally we modify the time-advanced values of the mass fractions $Y_m$ to be consistent with the corrected fluxes.

\subsection{Chemical mechanism and species transport properties}
\label{ssec:chem}

The chemical mechanism employed in this work combines the GRI3.0 \cite{SmithGRI:2000} for the oxidation of methane with the reaction mechanism for charged species reported in Belhi \emph{et al.} \cite{Belhi:2018}. The combined mechanism contains 61 species (not including electrons) and 386 reactions, and includes 10 ions (4 cations and 6 anions) as listed in Table~\ref{tab:ions}. In the remainder of the paper, charged species refers to the ions whereas charged particles also includes the  electron. Several studies have showed that anions are only present in very small quantities in freely evolving flames; electrons account for most of the negative charges. However, Belhi \emph{et al.} \cite{Belhi:2019} recently showed that including the anions (especially large anions such as CO$_3^-$ and HCO$_3^-$) is essential to reproduce the ionic wind motion observed experimentally.

\begin{table*}[ht!]
\centering
\begin{tabular}{| l | c c c c c c |}\hline
Cation & H$_3$O$^+$ & HCO$^+$ & C$_2$H$_3$O$^+$ & CH$_5$O$^+$ & & \\
$W_m$ [g/mol]& 19.02 & 29.02 & 43.05 & 33.05 & & \\\hline
Anions & OH$^-$ & O$^-$ & O$_2^-$ & CO$_3^-$ & HCO$_2^-$ & HCO$_3^-$ \\
$W_m$ [g/mol] & 17.01 & 16.00 & 32.00 & 60.01 & 44.01 & 61.02 \\\hline
\end{tabular}
\caption{List of ions included in the chemical mechanism along with their molecular weight.}
\label{tab:ions}
\end{table*}

The thermodynamic data for the charged species from the Burcat \cite{Burcat:2006} database were used. The computation of the transport properties for the charged particles listed in Table~\ref{tab:ions} uses the EGLIB library. Specific treatment of the ion/neutral or ion/ion collision is not investigated in this work, the use of (n,6,4) and Coulomb \cite{Mason:1988} interaction potentials for ions/neutrals and ions/ions collisions as described in Han \emph{et al.} \cite{Han:2015} will be studied in future work. 

The electron transport coefficients require a more detailed treatment. For low values of the reduced electric field $|\boldsymbol{E}|/\mathcal{N}$, where $\mathcal{N}$ is the background gas number density, the electrons are in thermal equilibrium with the mixture. In these conditions, the electron temperature is equal to that of the mixture: electrons are accelerated by the electric field $\boldsymbol{E}$, but the collision frequency with neutral species (represented by $\mathcal{N}$) is high enough to prevent the electrons from reaching high kinetic energy conditions. For higher values of $|\boldsymbol{E}|/\mathcal{N}$, the electrons gain sufficient kinetic energy that their energy (temperature) is higher than the remainder of the mixture. For this case the electrons are said to be non-thermal. Under these conditions, the evaluation of the electron transport coefficients require the computation of the evolution of the electron energy distribution function (EEDF) by solving the Boltzmann equation \cite{Bisetti:2014}. Additionally, the chemi-ionization reaction $\text{CH} + \text{O} \rightarrow \text{HCO}^+ + e^-$ is no longer the only chemical pathway producing electrons since impact ionization rates become important \cite{Bisetti:2014}. However, this last effect is not included in our framework at the present time and its relevance will be the subject of future studies.

\begin{figure*}[ht!]
\centering
\includegraphics[width=0.95\textwidth]{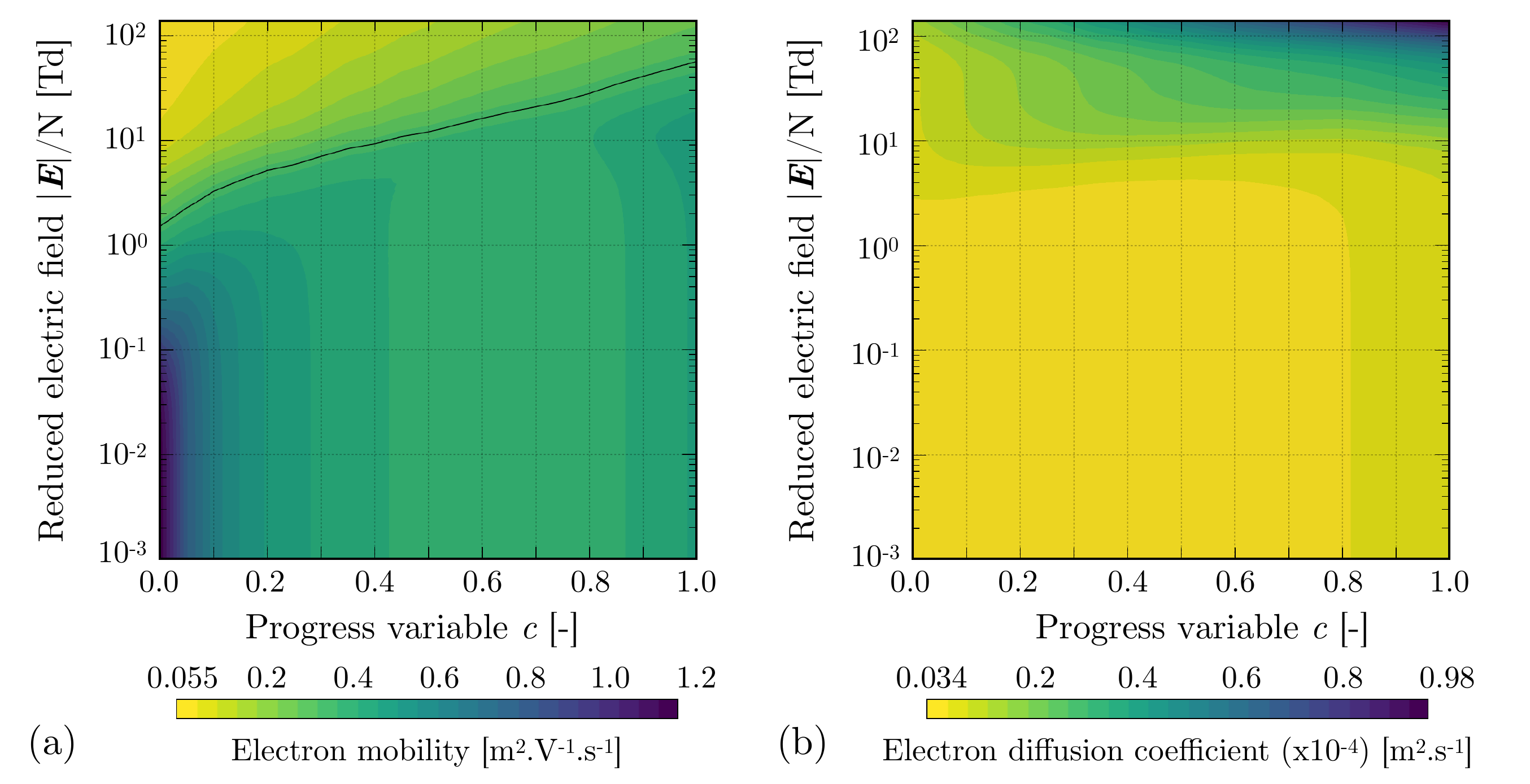}
\caption{(a) Electron mobility $\kappa_e$ as function of progress variable and reduced electric field. The full line corresponds to the constant value $\kappa_e = 0.4 $ m$^2$.V$^{-1}$.s$^{-1}$. (b) Electron diffusion coefficient $\kappa_e$ as function of progress variable and reduced electric field.}
\label{fig:electransp}
\end{figure*}

Previous studies employed a constant value of the electron mobility $\kappa_e = 0.4$ mJ$^{-1}$s$^{-1}$ \cite{Belhi:2017,Renzo:2018} since this constant value was found to provide a good agreement with simulations obtained from more detailed thermal electron transport calculations \cite{Bisetti:2012}. The framework developed in this work aims at simulating realistic engineering applications characterized by relatively high external voltages, conditions at which electrons can no longer be assumed thermal. In order to include a non-thermal electron transport coefficient without explicitly computing the evolution of the EEDF, the mixture composition and temperature are extracted from the simulation as function of the progress variable $c$:
\begin{equation}
c = \frac{T - T_{in}}{T_{max} - T_{in}}
\end{equation}
Note that other definitions of the progress variable could be used for fuels exhibiting more complex behaviors. This information is then used in the BOLSIG+ \cite{Hagelaar:2005} code to estimate the EEDF and the corresponding value of the electron mobility and diffusion coefficient at different values of $|\boldsymbol{E}|/\mathcal{N}$. The resulting two-dimensional tables are shown in Fig.~\ref{fig:electransp} and electron transport properties are extracted from these tables during the simulation using $c$ and $|\boldsymbol{E}|/\mathcal{N}$. Note that

\section{MISDC strategy}

\subsection{MISDC strategy}

The present strategy builds upon the MISDC methodology developed in Nonaka \emph{et al.} \cite{Nonaka:2012, Nonaka:2018}. As a brief reminder, the spectral deferred correction (SDC) method \cite{Dutt:2000} solves a system of ordinary differential equation:
\begin{align}
& \boldsymbol{\varphi}_t = \boldsymbol{F}(t,\boldsymbol{\varphi}(t)), \; \; t \in [t^n,t^{n+1}];\\
& \boldsymbol{\varphi}(t^n) = \boldsymbol{\varphi}^n
\end{align}
using the integral form:
\begin{equation}
\boldsymbol{\varphi}(t) = \boldsymbol{\varphi}^n + \int_{t^n}^t \boldsymbol{F}(\tau,\boldsymbol{\varphi}(\tau)) \; \text{d}\tau
\end{equation}
The SDC method generates successive approximation $\boldsymbol{\varphi}^{(k)}(t)$ of $\boldsymbol{\varphi}(t)$ using the update equation:
\begin{equation}
\boldsymbol{\varphi}^{(k+1)}(t) = \boldsymbol{\varphi}^n + \int_{t^n}^t \left[ \boldsymbol{F}(\boldsymbol{\varphi}^{(k+1)}) - \boldsymbol{F}(\boldsymbol{\varphi}^{(k)}) \right] \text{d}\tau + \int_{t^n}^t \boldsymbol{F}(\boldsymbol{\varphi}^{(k)}) \; \text{d}\tau
\end{equation}
where the explicit dependence of $\boldsymbol{F}$ and $\boldsymbol{\varphi}$ on $t$ in the integrals has been dropped for simplicity. By using a low-order approximation of the first integral and a more accurate quadrature rule for the second integral, the SDC method effectively constructs an arbitrary order (the order of the quadrature rule) solution by successive low-order corrections of the approximation $\boldsymbol{\varphi}^{(k)}(t)$. In MISDC \cite{Bourlioux:2003,Layton:2004}, $\boldsymbol{F}$ is decomposed into distinct processes, that can be treated separately in their own time scales:
\begin{equation}
\boldsymbol{F}(t,\boldsymbol{\varphi}(t)) \equiv \boldsymbol{A}(t,\boldsymbol{\varphi}(t)) + \boldsymbol{D}(t,\boldsymbol{\varphi}(t)) + \boldsymbol{R}(t,\boldsymbol{\varphi}(t))
\end{equation}
with $\boldsymbol{A}$, $\boldsymbol{D}$ and $\boldsymbol{R}$ referring here to the advection, diffusion and reaction processes, respectively. Here, following \cite{Nonaka:2012}, $\boldsymbol{A}(t,\boldsymbol{\varphi}(t))$ and $\boldsymbol{D}(t,\boldsymbol{\varphi}(t))$ are piece-wise constant over each time step. The former is evaluated using a second-order Godunov method while the latter is evaluated using a midpoint rule. To accommodate the stiffness of hydrocarbon chemical reactions, the update equation for the reaction is formulated as an ODE and integrated using a stiff ODE package such as CVODE. The effects of advection and diffusion are taken into account as temporally constant forcing terms in the chemical ODE integration (see \cite{Nonaka:2012} for more details on the integration procedure).

The main steps of the integration algorithm are summarized in Algorithm \ref{algo:misdc}. The set of transported thermodynamic scalars is written as $\boldsymbol{\varphi} = (\rho, \rho h, T, n_e, \rho Y_m)^T$. The superscript $n$ indicates the timestep and $(k)$ is the SDC iteration index. The diffusion operator for scalar $\varphi$ at time $t^n$ is written $\boldsymbol{D}_\varphi^{n}$, the $k$-th approximation of this operator at time $t^{n+1}$ is written $\boldsymbol{D}_\varphi^{n+1,(k)}$ and the $k$-th approximate of the advection operator obtained with the Godunov procedure is $\boldsymbol{A}^{n+1/2,(k)}_\varphi$.
The charged species drift flux appearing in Eq.~\ref{eq:diffusionfluxes} is non-symmetric and introduces numerical instabilities when discretized with the species diffusion flux using a second order centered scheme. To overcome this difficulty, the drift flux is treated along with the advective flux in the second-order Godunov procedure by constructing an effective velocity for each species, $m$:
\begin{equation}
\boldsymbol{U}^{(k)}_{ef,m} = \boldsymbol{U}^{(k)}_{adv} - \nu_m \kappa_m \boldsymbol{\nabla \phi}^{n+1,(k)}
\end{equation}

\begin{algorithm}
\SetAlgoLined
{\bf Get advection velocities: $\boldsymbol{U}_{adv}^{*}$} \\
{\bf Advance thermodynamic variables} \\
 \nonl (a) Initialize SDC predictor: \\
     \nonl  \qquad - Get scalar transport properties from $\boldsymbol{\varphi}^{n}$\\
     \nonl  \qquad - Solve Poisson equation for $\phi^n$\\
     \nonl  \qquad - Compute initial diffusion operators $\boldsymbol{D}_\varphi^{n}$\\
     \nonl  \qquad - Initialize SDC predictor: $\boldsymbol{\varphi}^{n+1,(0)} \leftarrow  \boldsymbol{\varphi}^{n}$, $\boldsymbol{D}_\varphi^{n+1,(0)} \leftarrow \boldsymbol{D}_\varphi^{n}$\\
 \nonl (b) {\bf do} $k$ : 1, $k_{max}$ ($>$ 2 for 2$^{nd}$-order)\\
     \nonl  \qquad - Get scalar transport properties from $\boldsymbol{\varphi}^{n+1,(k)}$\\
     \nonl  \qquad - Compute approximate diffusion operators $\boldsymbol{D}_\varphi^{n+1,(k)}$\\
     \nonl  \qquad - Compute projected advection velocity $\boldsymbol{U}^{(k)}_{adv}$\\
     \nonl  \qquad - Compute explicit advection operators $\boldsymbol{A}^{n+1/2,(k)}_\varphi$: \\
     \nonl  \qquad \qquad $\cdot$ Compute effective velocity for each species $\boldsymbol{U}^{(k)}_{ef,m} = \boldsymbol{U}^{(k)}_{adv} - \nu_m \kappa_m \boldsymbol{\nabla \phi}^{n+1,(k)}$\\
     \nonl  \qquad \qquad $\cdot$ Use second-order Godunov to get advection fluxes $(\varphi \boldsymbol{U}_{ef,\varphi} )^{n+1/2,{(k)}}$ \\
     \nonl  \qquad - Compute implicitly the species and enthalpy diffusion $\boldsymbol{D}^{n+1,(k+1)}_{\varphi,AD}$\\
     \nonl  \qquad - Solve implicit non-linear electron/Poisson system \\
     \nonl  \qquad \qquad $\cdot$ Compute provisional charged species fields $\widetilde{\rho Y}^{n+1,(k+1)}_m$ \\
     \nonl  \qquad \qquad $\cdot$ Use algorithm from Section \ref{ssec:pnp} to get $n_e^{n+1,(k+1)}$ and $\phi^{n+1,(k+1)}$ \\
     \nonl  \qquad - Integrate species reaction and enthalpy evolution over $\Delta t$ and evaluate the reaction term \\
     \nonl  \qquad   $\boldsymbol{I}^{(k+1)}_{R,\varphi}$\\     
{\bf Advance velocity} \\
\caption{Time step: $t^n \rightarrow t^{n+1}$}
\label{algo:misdc}
\end{algorithm}

The resolution of the coupled electron/electrostatic potential non-linear system requires provisional charged species mass fraction $\widetilde{\rho Y}^{n+1,(k+1)}_m$:
 \begin{equation}
\widetilde{\rho Y}^{n+1,k+1}_m = \rho Y^{n}  + \Delta t \left[ \boldsymbol{A}^{n+1/2,(k)}_m + \frac{1}{2}\left(\boldsymbol{D}^{n}_m - \boldsymbol{D}^{n+1,(k)}_m \right) + \boldsymbol{D}^{n+1,(k+1)}_{m,AD} + \boldsymbol{I}^{(k)}_{R,m} \right ]
 \end{equation}
 where $\boldsymbol{I}^{(k)}_{R,m}$ is the integrated representation of the reaction term for species $m$ from the previous SDC iteration.

\subsection{Non-linear implicit solution}
\label{ssec:pnp}

At each SDC iteration, we solve the non-linear system formed by the electron conservation equation (Eq. (\ref{eq:elec})) and the electrostatic potential equation (Eq. (\ref{eq:poisson})):
\begin{subequations}
   \begin{empheq}[left=\empheqlbrace]{align}
    & \dfrac{\partial (n_e)}{\partial t} = - \boldsymbol{\nabla} \cdot n_e(\Uvec - \kappa_e\boldsymbol{\nabla \phi}) + \boldsymbol{\nabla}\cdot D_e\boldsymbol{\nabla n_e} + I_{R,e} \\
    & \varepsilon_0\varepsilon_r \boldsymbol{\nabla}^2\phi  = - \sum_m z_m \widetilde{\rho Y}_m + e n_e,
\end{empheq}\label{eq:pnp}
\end{subequations}
 where $\widetilde{\rho Y}^{n+1,k+1}_m$ is the provisional charged species mass fraction at the current SDC iteration and $I_{R,e}$ is the last evaluation of the electron chemical source term. 
 
Using a first order backward Euler time discretization, the implicit non-linear system can be written as:
\begin{subequations}
   \begin{empheq}[left=\empheqlbrace]{align}
    & - n_e^{n+1} - \Delta t \boldsymbol{\nabla} \cdot n_e^{n+1}(\Uvec - \kappa_e\boldsymbol{\nabla \phi}^{n+1}) - \Delta t \boldsymbol{\nabla}\cdot D_e\boldsymbol{\nabla n_e}^{n+1} + f_e = 0 \\
    & - n_e^{n+1} + \frac{\varepsilon_0\varepsilon_r}{e} \boldsymbol{\nabla}^2\phi^{n+1} + f_{\phi} = 0,
\end{empheq}\label{eq:nls}
\end{subequations}
where $f_e = - I_{R,e}^n + n_e^n$ and $f_{\phi} = \sum_m z_m \widetilde{\rho Y}^{n+1,k+1}_m /e$. Introducing $\boldsymbol{X} = (\boldsymbol{n_e}, \boldsymbol{\phi})$, Eq.~(\ref{eq:nls}) can be written as $F(\boldsymbol{X}) = 0$, where $F(\boldsymbol{X})$ is the non-linear residual. This system is solved using a Jacobian-free Newton-Krylov (JFNK) method \cite{Knoll:2004}.

The basis of JFNK is the iterative non-linear Newton solution, where at each iteration $l$, a linear system of the form:
\begin{equation}
\mathcal{J}^{(l)}\delta\boldsymbol{X}^{(l)} = -F(\boldsymbol{X}^{(l)})
\label{eq:newton_ite}
\end{equation}
is solved. Here, $\delta\boldsymbol{X}^{(l)} = \boldsymbol{X}^{(l+1)} - \boldsymbol{X}^{(l)}$ is the Newton update and $\mathcal{J}^{(l)} = \mathcal{J}(\boldsymbol{X}^{(l)}) = \partial F(\boldsymbol{X}^{(l)})/\partial \boldsymbol{X}$ is the system Jacobian matrix. In practice, the components of $\boldsymbol{X}^{(l)}$ and $F(\boldsymbol{X}^{(l)})$ can have entries than span a large range of values which can affect the solution of the linear system (\ref{eq:newton_ite}) and destroy the convergence properties of Newton's method. To address this issue, Eq.~(\ref{eq:newton_ite}) is scaled by two diagonal matrices $S_F$ and $S_X$:
\begin{equation}
(S_F^{-1}\mathcal{J}^{(l)}S_X)(S_X^{-1}\delta\boldsymbol{X}^{(l)}) = -S_F^{-1}F(\boldsymbol{X}^{(l)})
\label{eq:newton_sc}
\end{equation}
where $S_F$ contains typical values of $F(n_e)$ and $F(\phi)$ respectively, and $S_X$ contains typical values of $n_e$ and $\phi$. The typical values are evaluated at the beginning of the non-linear iterations since the apropriate values may evolve with the solution.
The Newton iterations are stopped when the norm of scaled residual is reduced by $\epsilon_{F}$ orders of magnitude or the scaled magnitude of the Newton step drops below a certain value $\epsilon_X$:
\begin{align}
||S_F^{-1}F(\boldsymbol{X}^{(l)})||_{\infty} & < \epsilon_F \\
||S_X^{-1}\delta\boldsymbol{X}^{(l)}||_{\infty} & < \epsilon_X
\end{align}
These tolerances must be chosen to ensure that the non-linear solution residual remains smaller than the truncation error of the numerical schemes. A backtracking linesearch algorithm is employed for globalization of the Newton method \cite{Dennis:1996}. 

For large non-linear systems encountered in multi-dimensional simulations, the computational cost and memory requirements of solving a linear system with a direct solver at each Newton iteration are prohibitive. Our implementation thus uses the GMRES Krylov method \cite{Saad:1986} to solve the scaled linear system (\ref{eq:newton_sc}). For clarity, the left and right scaling matrices will not be carried in the following description and the outer (Newton) iteration index $l$ is dropped (the scaling is implemented in the code; however). The GMRES starts with an initial guess $\delta\boldsymbol{X}_0$, and the corresponding residual $\boldsymbol{r_0} = - F(\boldsymbol{X}_0) - \mathcal{J}\delta\boldsymbol{X}_0$. In the context of a Newton-Krylov method, $\delta\boldsymbol{X}_0 = 0$ is used since the Newton step tends toward zero as we go through the Newton iterations. At the $p^{th}$ iteration of the GMRES method, we construct an approximation $\delta\boldsymbol{X}_p$ of the solution by solving a minimization problem in the Krylov subspace $\mathcal{K}_p$ of $\mathcal{J}$:
\begin{equation}
\mathcal{K}_p(\mathcal{J},\boldsymbol{r}_0) = \text{span}(\boldsymbol{r}_0, \mathcal{J}\boldsymbol{r}_0, \mathcal{J}^2\boldsymbol{r}_0,..., \mathcal{J}^{p-1}\boldsymbol{r}_0)
\end{equation}
It can be seen that the GMRES method only needs the action of the Jacobian matrix on a vector. For large linear systems, the construction and storage of matrix $\mathcal{J}$ can hinder the performance and the scalability of the algorithm. In the JFNK context, the explicit construction of $\mathcal{J}$ is dropped in favor of a finite difference approximation of the the matrix/vector product $\mathcal{J} \boldsymbol{v}$:
\begin{equation}
\mathcal{J}(\boldsymbol{X}) \boldsymbol{v} = \frac{F(\boldsymbol{X} + \varepsilon_{FD} \boldsymbol{v}) - F(\boldsymbol{X})}{\varepsilon_{FD}}
\label{eq:jf_dfapprox}
\end{equation}
where $\varepsilon_{FD}$ is a small number. The quality of the approximation of $\mathcal{J}.\boldsymbol{v}$ depends on the choice of $\varepsilon_{FD}$. Here we use the method employed in the Trilinos package \cite{Heroux:2005}:
\begin{equation}
\varepsilon_{FD} = \lambda_{FD} \left( \lambda_{FD} + \frac{|\boldsymbol{X}|}{|\boldsymbol{v}|} \right)
\end{equation}
where $\lambda_{FD} = \varepsilon_{mach}^{1/3}$ is a small parameter related to the machine precision $\varepsilon_{mach}$. The linear solver is iterated until:
\begin{equation}
||\mathcal{J}\delta\boldsymbol{X}_p+F(\boldsymbol{X})||_2 < \gamma ||F(\boldsymbol{X})||_2
\end{equation}
A constant value of $\gamma$ is kept throughout the simulation and the effect of the choice of $\gamma$ on the non-linear system solution will be assessed in Section \ref{ssec:algo_perf}.

The performance of the JFNK depends strongly on the number of GMRES iterations required to solve (\ref{eq:newton_sc}). If $\mathcal{J}$ has a large condition number, the Krylov method requires a large number of iterations to converge. In this case, it is necessary to apply a preconditioner to the linear system:
\begin{equation}
P^{-1}\mathcal{J}\delta\boldsymbol{X} = -P^{-1}F(\boldsymbol{X})
\end{equation}
where $P$ is an approximation of $\mathcal{J}$, such that $P^{-1}\mathcal{J} \sim \mathcal{I}$. The main objective of the preconditioner is to cluster the eigenvalues of the resulting $P^{-1}\mathcal{J}$ matrix, allowing the GMRES method to find a good $\delta\boldsymbol{X}_p$ in a small Krylov space (i.e. small number of iterations).
To construct the preconditioner, we start by linearizing Eq.~(\ref{eq:nls}):
\begin{subequations}
   \begin{empheq}[left=\empheqlbrace]{align}
    & - \delta n_e^{n+1} + \Delta t \underbrace{[\boldsymbol{\nabla}\cdot D_e\boldsymbol{\nabla} -\boldsymbol{\nabla} \cdot (\Uvec - \kappa_e\boldsymbol{\nabla \phi}^{n+1})]}_{\mathcal{D}_f} \delta n_e^{n+1} + \Delta t \underbrace{\boldsymbol{\nabla} \cdot n_e^{n+1} \kappa_e\boldsymbol{\nabla}}_{\mathcal{D}_r} \delta  \phi^{n+1} = 0 \\
    & - \delta n_e^{n+1} + \underbrace{\frac{\varepsilon_0\varepsilon_r}{e} \boldsymbol{\nabla}^2}_{\mathcal{L}}\delta\phi^{n+1} = 0,
\end{empheq}\label{eq:linear_nls}
\end{subequations}
This allows us to write the block matrix form of the Jacobian $\mathcal{J}$ resulting from the spatio-temporal discretization of Eq.~(\ref{eq:nls}):
\begin{equation}
\mathcal{J} = \begin{pmatrix} (\Delta t \boldsymbol{\mathcal{D}_f} - \boldsymbol{\mathcal{I}}) & \Delta t \boldsymbol{\mathcal{D}_r} \\ \boldsymbol{\mathcal{I}_e} & \boldsymbol{\mathcal{L}} \end{pmatrix}
\end{equation}
where the block matrices $\boldsymbol{\mathcal{D}_f}$, $\boldsymbol{\mathcal{D}_r}$ and $\boldsymbol{\mathcal{L}}$ are the spatial operators underlined in Eq.~\ref{eq:linear_nls}. Note that $\boldsymbol{\mathcal{I}_e}$ actually differs from the identity matrix because of the scaling applied to the linear system (\ref{eq:newton_sc}).
Schur factorization of the inverse of the 2 $\times$ 2 block Jacobian is written as:
\begin{equation}
P^{-1} = \mathcal{J}^{-1} = \begin{pmatrix} \boldsymbol{\mathcal{I}} & - (\Delta t \boldsymbol{\mathcal{D}_f} - \boldsymbol{\mathcal{I}})^{-1} \Delta t \boldsymbol{\mathcal{D}_r} \\ 0 & \boldsymbol{\mathcal{I}} \end{pmatrix} \begin{pmatrix} (\Delta t \boldsymbol{\mathcal{D}_f} - \boldsymbol{\mathcal{I}})^{-1} & 0 \\ 0 & \boldsymbol{\mathcal{S}}^{-1} \end{pmatrix} \begin{pmatrix} \boldsymbol{\mathcal{I}} & 0 \\ - \boldsymbol{\mathcal{I}_e} (\Delta t \boldsymbol{\mathcal{D}_f} - \boldsymbol{\mathcal{I}})^{-1} & \boldsymbol{\mathcal{I}} \end{pmatrix}
\label{eq:prec}
\end{equation}
where $\boldsymbol{\mathcal{S}} = \boldsymbol{\mathcal{L}} - \boldsymbol{\mathcal{I}_e} (\Delta t \boldsymbol{\mathcal{D}_f} - \boldsymbol{\mathcal{I}})^{-1} \Delta t \boldsymbol{\mathcal{D}_r}$ is the Schur complement of $\mathcal{J}$. Here, $P^{-1}$ is the exact inverse of the Jacobian matrix and it only requires $(\Delta t \boldsymbol{\mathcal{D}_f} - \boldsymbol{\mathcal{I}})^{-1}$ and $\boldsymbol{\mathcal{S}}^{-1}$, both of which are easier to invert than $\mathcal{J}$. However, computing $\boldsymbol{\mathcal{S}}^{-1}$ is still difficult since the construction of $\boldsymbol{\mathcal{S}}$ requires the solution of $(\Delta t \boldsymbol{\mathcal{D}_f} - \boldsymbol{\mathcal{I}})^{-1}$. Instead, we use an approximation $\widetilde{\boldsymbol{\mathcal{S}}} = \boldsymbol{\mathcal{L}} + \boldsymbol{\mathcal{I}_e} \Delta t \boldsymbol{\mathcal{D}_r}$ of $\boldsymbol{\mathcal{S}}$ that is easier to solve. It can be seen that for small time steps, $\widetilde{\boldsymbol{\mathcal{S}}}$ is a good approximation of $\boldsymbol{\mathcal{S}}$. Both $(\Delta t \boldsymbol{\mathcal{D}_f} - \boldsymbol{\mathcal{I}})$ and $\widetilde{\boldsymbol{\mathcal{S}}}$ are then diagonally dominant and can be solved effectively using a multi-grid (MG) approach. The present implementation uses a standard V-cycle approach with red-black Gauss-Siedel relaxation to solve both linear systems to a tolerance $\gamma_{MG}$. The effect of $\gamma_{MG}$ on the performance of the JFNK is evaluated in Section~\ref{ssec:algo_perf}. Applying $P^{-1}$ to any vector $\boldsymbol{v}$ requires the application of the successive matrices of Eq.~(\ref{eq:prec}) to $\boldsymbol{v}$. In its classical Schur factorization form (\ref{eq:prec}), this entails four MG solves (three solves of $(\Delta t \boldsymbol{\mathcal{D}_f} - \boldsymbol{\mathcal{I}})^{-1}$ and one of $\widetilde{\boldsymbol{\mathcal{S}}}^{-1}$). To save one MG solve, the block factorization (\ref{eq:prec}) is rewritten in the following form:
\begin{equation}
\widetilde{P}^{-1} = \begin{pmatrix} \boldsymbol{\mathcal{I}} & - (\Delta t \boldsymbol{\mathcal{D}_f} - \boldsymbol{\mathcal{I}})^{-1} \Delta t \boldsymbol{\mathcal{D}_r} \\ 0 & \boldsymbol{\mathcal{I}} \end{pmatrix} \begin{pmatrix} \boldsymbol{\mathcal{I}} & 0 \\ 0 & \widetilde{\boldsymbol{\mathcal{S}}}^{-1} \end{pmatrix} \begin{pmatrix} \boldsymbol{\mathcal{I}} & 0 \\ \boldsymbol{\mathcal{I}}_e & \boldsymbol{\mathcal{I}} \end{pmatrix}  \begin{pmatrix} (\Delta t \boldsymbol{\mathcal{D}_f} - \boldsymbol{\mathcal{I}})^{-1} & 0 \\ 0 & \boldsymbol{\mathcal{I}} \end{pmatrix}
\label{eq:prec_recast}
\end{equation}

The solution of the implicit non-linear system is summarized in Algorithm \ref{algo:jfnk}. Superscript $l$ corresponds to the Newton iteration index while subscript $p$ is the GMRES iteration index.

\begin{algorithm}
\SetAlgoLined
{Get typical values of $\boldsymbol{X}^{(0)}$ and $F(\boldsymbol{X}^{(0)})$ to fill $S_X$ and $S_F$}\\
{\bf do while} : $||S_F^{-1}F(\boldsymbol{X}^{(l)})||_{\infty} >  \epsilon_F$ \\
 \nonl \qquad - Build MG operators for $P^{-1}$ with current $\boldsymbol{X}^{(l)}$\\
 \nonl \qquad - $\delta\boldsymbol{X}_0 = 0$ \\
 \nonl \qquad - $\boldsymbol{r}_0 = P^{-1}F(\boldsymbol{X}^{(l)})$ \\
 \nonl \qquad - Initialize Krylov subspace base vector $\boldsymbol{K}_0 = \boldsymbol{r}_0/||\boldsymbol{r}_0||$ \\
 \nonl \qquad {\bf do while} : $||\mathcal{J}^{(l)}\delta\boldsymbol{X}_p+F(\boldsymbol{X}^{(l)})||_2 > \gamma ||F(\boldsymbol{X}^{(l)})||_2$ \\
     \nonl  \qquad \qquad - Compute $\boldsymbol{K}_p = \widetilde{P}^{-1} \mathcal{J}^{(l)}\boldsymbol{K}_{p-1}$\\
         \nonl  \qquad \qquad \qquad - FD approximation of the matrix-vector product $\mathcal{J}^{(l)}\boldsymbol{K}_{p-1}$ (Eq. (\ref{eq:jf_dfapprox}))\\
         \nonl  \qquad \qquad \qquad - Apply Eq.(\ref{eq:prec_recast}), using MG to solve the $(\Delta t \boldsymbol{\mathcal{D}_f} - \boldsymbol{\mathcal{I}})^{-1}$ and $\widetilde{\boldsymbol{\mathcal{S}}}^{-1}$ blocks\\ 
     \nonl  \qquad \qquad - Gram-Schmidt method to orthogonalize $\boldsymbol{K}_p$\\
     \nonl  \qquad \qquad - Find $\delta\boldsymbol{X}_p$ that minimizes residual $\boldsymbol{r}_p$ in the Krylov subspace $\mathcal{K}_p(P^{-1}\mathcal{J}^{(l)},\boldsymbol{r}_0)$\\
\nonl \qquad - Evaluate $\lambda$ such that $||F(\boldsymbol{X}^{(l)} + \lambda \delta\boldsymbol{X}_p)|| <  ||F(\boldsymbol{X}^{(l)}) - \alpha \lambda \nabla F(\boldsymbol{X}^{(l)})' \delta\boldsymbol{X}_p||$\\
\nonl \qquad - Update $\boldsymbol{X}^{(l+1)} = \boldsymbol{X}^{(l)} + \lambda \delta\boldsymbol{X}_p$\\
\caption{JFNK resolution}
\label{algo:jfnk}
\end{algorithm}

\section{Numerical experiments}

We first evaluate the robustness and performance of the proposed algorithm in order to optimize the numerical parameters and tolerances employed in the JFNK. Then, simulations of steady one-dimensional premixed methane/air flames subject to DC electric fields are performed in order to estimate the accuracy of the complete algorithm and provide comparisons with experimental data. Finally, the behavior of flames subjected to AC electric fields at various frequencies is analyzed.

\subsection{Numerical set-up}

Throughout this section, we consider an unstrained one-dimensional burner-stabilized premixed methane/air flame. The operating conditions correspond to the experimental study of Speelman \emph{et al.} \cite{Speelman:2015}: the inlet velocity is set to the flame speed of a stoichiometric methane/air flame at T = 300 K while the inlet temperature is set to T = 350 K, such that the flame is stabilized on the left boundary of the domain. Simulations are initialized from a resolved CANTERA \cite{Cantera:2017} solution ($\sim$ 4000 unequally-spaced grid points), that does not include the effect of the electric field. The CANTERA solution is interpolated onto a set of uniform grids with varying resolution, and simulations are evolved initially without external electric forcing for 5 ms in order to eliminate any spurious artifacts introduced by the initialization. Subsequently, the external electric field is activated and set to the desired values. The main characteristics of the simulations are listed in Table \ref{tab:prem_charac}. Unless otherwise specified, the numerical parameters ( $k_{max}$, $\gamma$, ...) listed in Table \ref{tab:prem_charac} are employed.

\begin{table*}[ht!]
\centering
\begin{tabular}{| c c c c c c |}
\hline
 \multicolumn{6}{| c |}{Operating conditions}\\\hline
 T$_{in}$ [K] & $U_{adv,in}$ [m/s] & Pressure [Pa] & Y$_{fuel,in}$ & Y$_{O2,in}$  & Y$_{N2,in}$\\
 350.0 & 0.371 & 101325.0 & 0.055 & 0.220 & 0.725  \\\hline\hline
\multicolumn{6}{| c |}{Numerical parameters}\\\hline
L [m] & $n_x$ & $\Delta_x$ [$\mu$m] & $k_{max}$ & $\gamma$ & $\gamma_{MG}$  \\
0.01 & [128,2048] & [156,9.77] & 4 &  1.0e$^{-4}$ & 1.0e$^{-4}$ \\\hline
\end{tabular}
\caption{Characteristic of the 1D laminar premixed flame}
\label{tab:prem_charac}
\end{table*}

\begin{figure*}[ht!]
\centering
\includegraphics[width=0.5\textwidth]{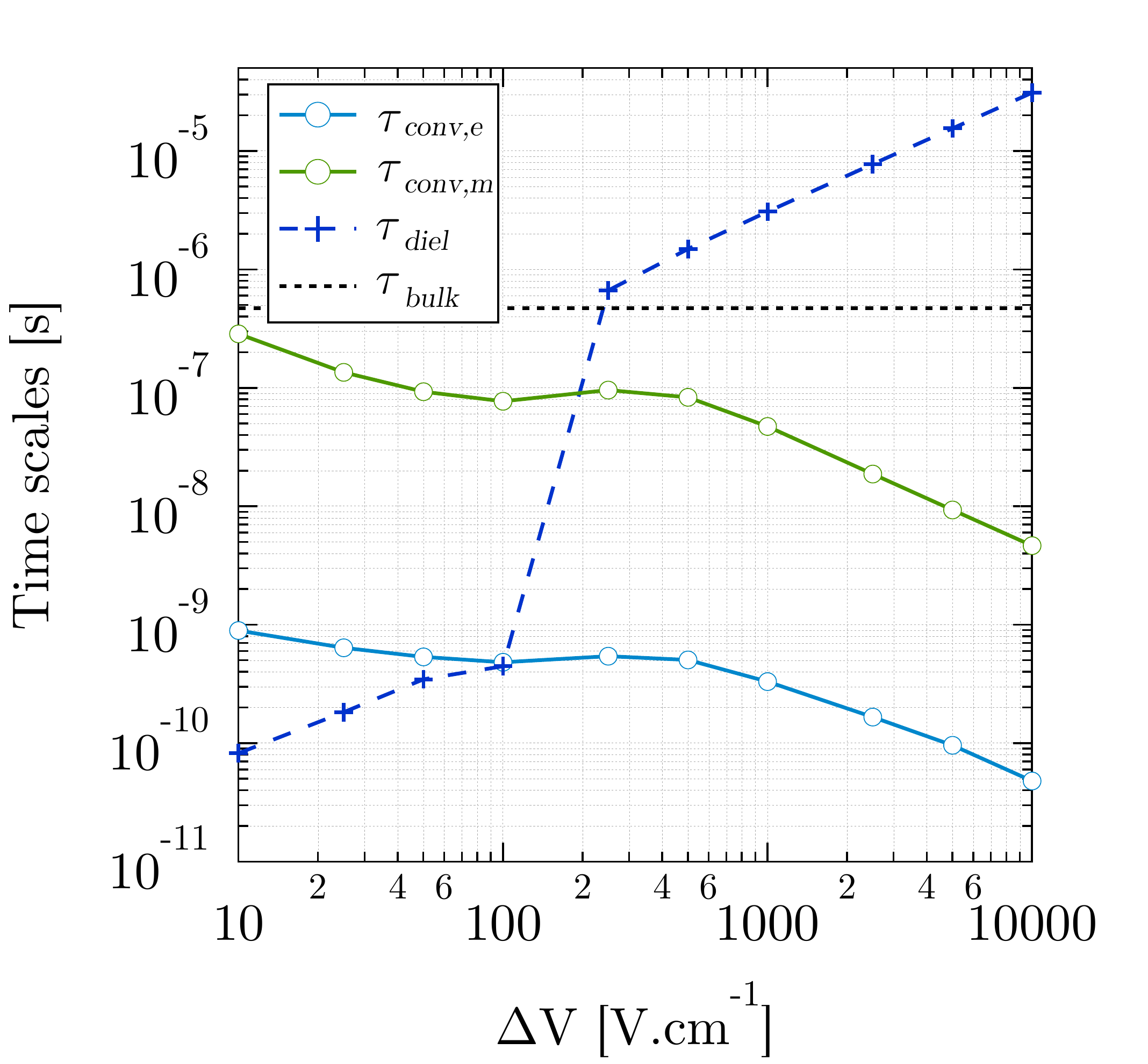}
\caption{Evolution of the typical value of the simulation time scales against the external voltage $\Delta V$ for a $n_x = 512$ mesh.}
\label{fig:timestep}
\end{figure*}

The interactions of the electric field with the charged particles in the flame introduces additional time scales compared to classical reactive flow simulations. The following is an overview of the relevant characteristic time scales and summarizes the specific treatment used here in the numerical strategy: 
\begin{itemize}
\item bulk advective time scale:
\begin{equation}
\tau_{bulk} = \frac{\Delta_x}{U_{adv}}
\end{equation}
\item species/electron diffusive time scale:
\begin{equation}
\tau_{diff,m} = \frac{\Delta_x^2}{2\;d\;\max_{\{m \in N_p\}}(D_m)}
\end{equation}
where $d$ is the number of dimensions.
\item chemical reaction time scale:
\begin{equation}
\tau_{chem} = \frac{\rho}{\max_{\{m \in N_p\}}(\dot{\omega}_m)}
\end{equation}
where $\dot{\omega}_m$ is chemical production rate per volume of species $m$.
\item charged species/electron effective convective time scale:
\begin{equation}
\tau_{conv,m} = \frac{\Delta_x}{\max_{\{m \in N_c\}}(U_{adv} + \kappa_m E)}
\end{equation}
\begin{equation}
\tau_{conv,e} = \frac{\Delta_x}{U_{adv} + \kappa_e E}
\end{equation}
where the drift velocity of the species is considered. Note that for large values of the external electric field, the drift velocity can oppose the convective velocity and its magnitude can exceed it. Additionally, the large mobility of the electrons results in a more stringent time step constraint, compared to ions.
\item electron dielectric relaxation time scale characterizes the response of the electric field to a change in the electron distribution:
\begin{equation}
\tau_{diel} = \frac{\epsilon_0 \epsilon_r}{e \kappa_e n_e}
\end{equation}
\end{itemize}
The first three time scales are common in reactive flow simulations. In most combustion simulations using detailed chemical kinetics, the chemical time-step constraint is alleviated in the numerical implementation by using a stiff ODE integrator. Additionally, we use a semi-implicit Crank-Nicholson method for conduction and species diffusion which enables time steps larger than the fast diffusive time scales of light species. The advection of charged species is treated time explicitly so that the advective time scale constrains the overall simulation time-step. Although this often results in time steps smaller than $\tau_{bulk}$, the charged species time scales are still several orders of magnitude larger than that of the electrons. Typical values of the various time scales are plotted against the external electric forcing $\Delta V$ in Fig. \ref{fig:timestep}. The data is based on a $n_x = 512$ grid points simulation, corresponding to $\Delta_x = 19.5 \mu$m. The bulk advective time scale is only shown as a reference for the classical low Mach number time constraint. Both $\tau_{conv,e}$ and $\tau_{conv,m}$ decrease with increasing values of $\Delta V$; $\tau_{conv,e}$ is approximately four orders of magnitude smaller than $\tau_{bulk}$. Both advective time scales exhibit a plateau at around $\Delta V = 250$ V.cm$^{-1}$, corresponding to the saturation voltage. At the same location, the dielectric time scale jumps to exceed $\tau_{bulk}$. This is behavior is related to the drop in peak electron number density as the external voltage exceeds the saturation value. Across the range of $\Delta V$ considered, the time scales associated with electrons are several orders of magnitude more stringent than the others, thus highlighting the need for an implicit treatment of the electrons.

\subsection{Iterative solvers performance}
\label{ssec:algo_perf}

Solution of the implicit non-linear electron/electrostatic potential system with JFNK involves several tolerances, which can have a significant impact on both the robustness and performance of the proposed methodology. A series of tests are performed in order to evaluate the optimal settings.

\begin{figure*}[ht!]
\centering
\includegraphics[width=0.9\textwidth]{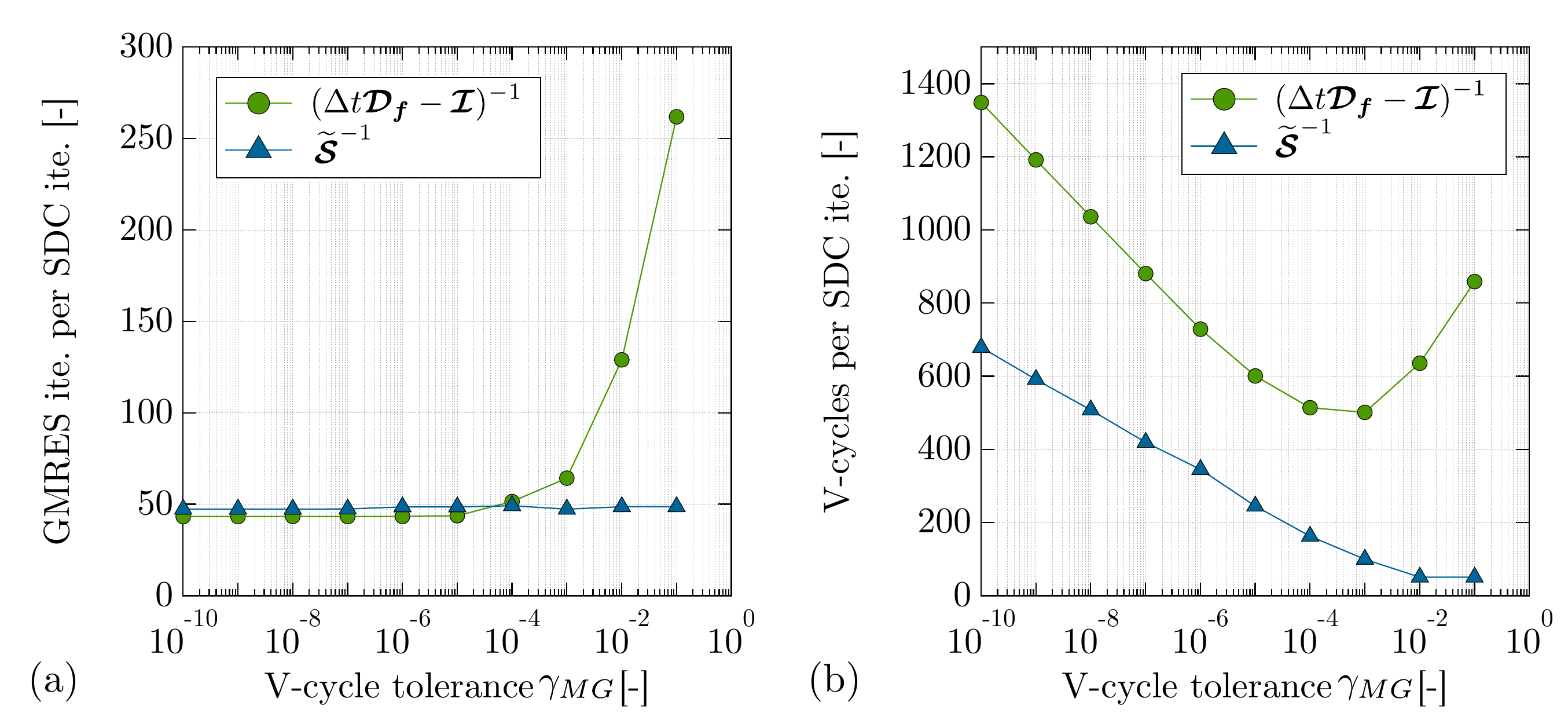}
\caption{(a) Total number of GMRES iterations per SDC iterations as function of the MG tolerance $\gamma_{MG}$. (b) Total number of V-cycle per SDC iterations as function of $\gamma_{MG}$.}
\label{fig:MG_tests}
\end{figure*}

At the lower level of the JFNK algorithm is the application of the inverse of the preconditioner $\widetilde{P}$ on the GMRES basis vectors (see Eq.~(\ref{eq:prec_recast})). This requires three MG solutions (two solutions of $(\Delta t \boldsymbol{\mathcal{D}_f} - \mathcal{I})^{-1}$ and one solution of $\widetilde{\boldsymbol{\mathcal{S}}}^{-1}$) using a standard V-cycle approach \cite{Briggs:2000}. Two relaxation operations are applied going down and up each level of the V-cycle based on a Red-Black Gauss Siedel. Figure~\ref{fig:MG_tests}(a) shows the total number of GMRES iterations per SDC iteration, as function of the V-cycle tolerance $\gamma_{MG}$. Figure~\ref{fig:MG_tests}(b) shows the total number of V-cycle as function of $\gamma_{MG}$. To separate the effect of each block on the performance of the preconditioner, the MG tolerance is tested for one block while the other is solved exactly (using a tri-diagonal solver in the present one-dimensional case). The number of GMRES iterations is only marginally affected by the multigrid tolerance on $\widetilde{\boldsymbol{\mathcal{S}}}^{-1}$, while loose tolerance on $(\Delta t \boldsymbol{\mathcal{D}_f} - \mathcal{I})^{-1}$ results in a large increase of the number of iterations. However, the number of V-cycles directly increases the CPU cost of the algorithm and, in the present cases, the trade-off between the MG tolerance and the total number of V-cycles shows V-cycles minimized around $\gamma_{MG} \sim 1.0e^{-4}$ as can be observed in Figure~\ref{fig:MG_tests}(b).

\begin{figure*}[ht!]
\centering
\includegraphics[width=0.55\textwidth]{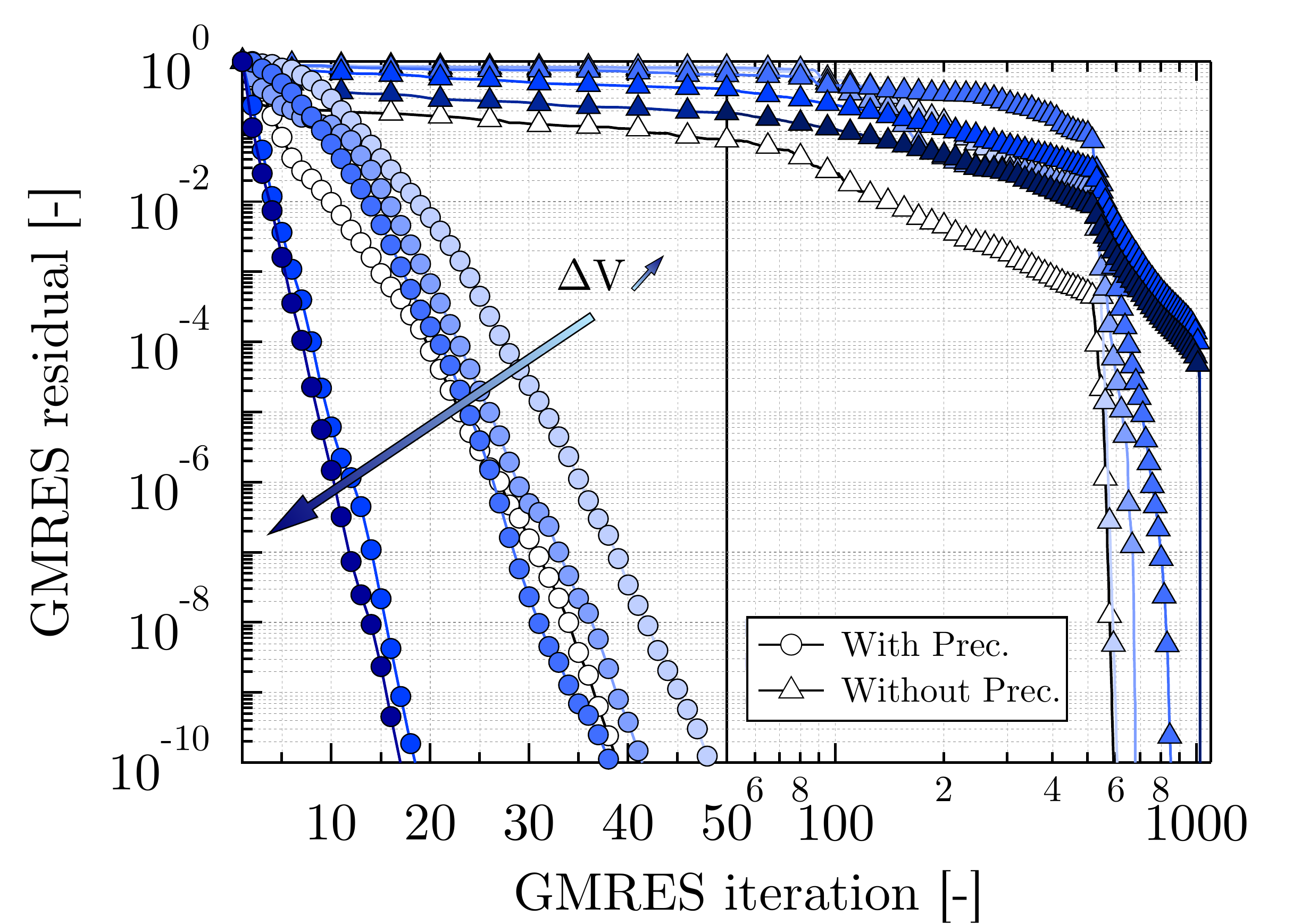}
\caption{Convergence of the normalized GMRES residual with and without preconditioner in the $n_x$ = 512 case and $\Delta V = 0, 10, 50, 100, 500, 1000$ V.}
\label{fig:Precond_tests}
\end{figure*}

Given these settings, the efficiency of the preconditioner can be directly evaluated by comparing the convergence of the GMRES solver with and without preconditioning. Figure~\ref{fig:Precond_tests} shows the GMRES residual as function of the GMRES iteration count for different values of $\Delta V$ both with and without the preconditioner. The preconditioned systems converge 20 to 50 times faster, regardless of the operating conditions. Note that simulations are performed at a constant CFL$_{conv,m}$, so that the time step is reduced as $\Delta V$ increases, resulting in a more efficient preconditioning ($\widetilde{\boldsymbol{\mathcal{S}}}^{-1}$ tends towards $\boldsymbol{\mathcal{S}}^{-1}$ as $\Delta_t$ decreases). Additionally, the preconditioned system convergence is only marginally affected by the size of the system whereas the unpreconditioned system convergence rate decreases with $n_x$ (not shown).

\subsection{Method convergence}
\label{ssec:algo_conv}

\begin{figure*}[ht!]
\centering
\includegraphics[width=0.95\textwidth]{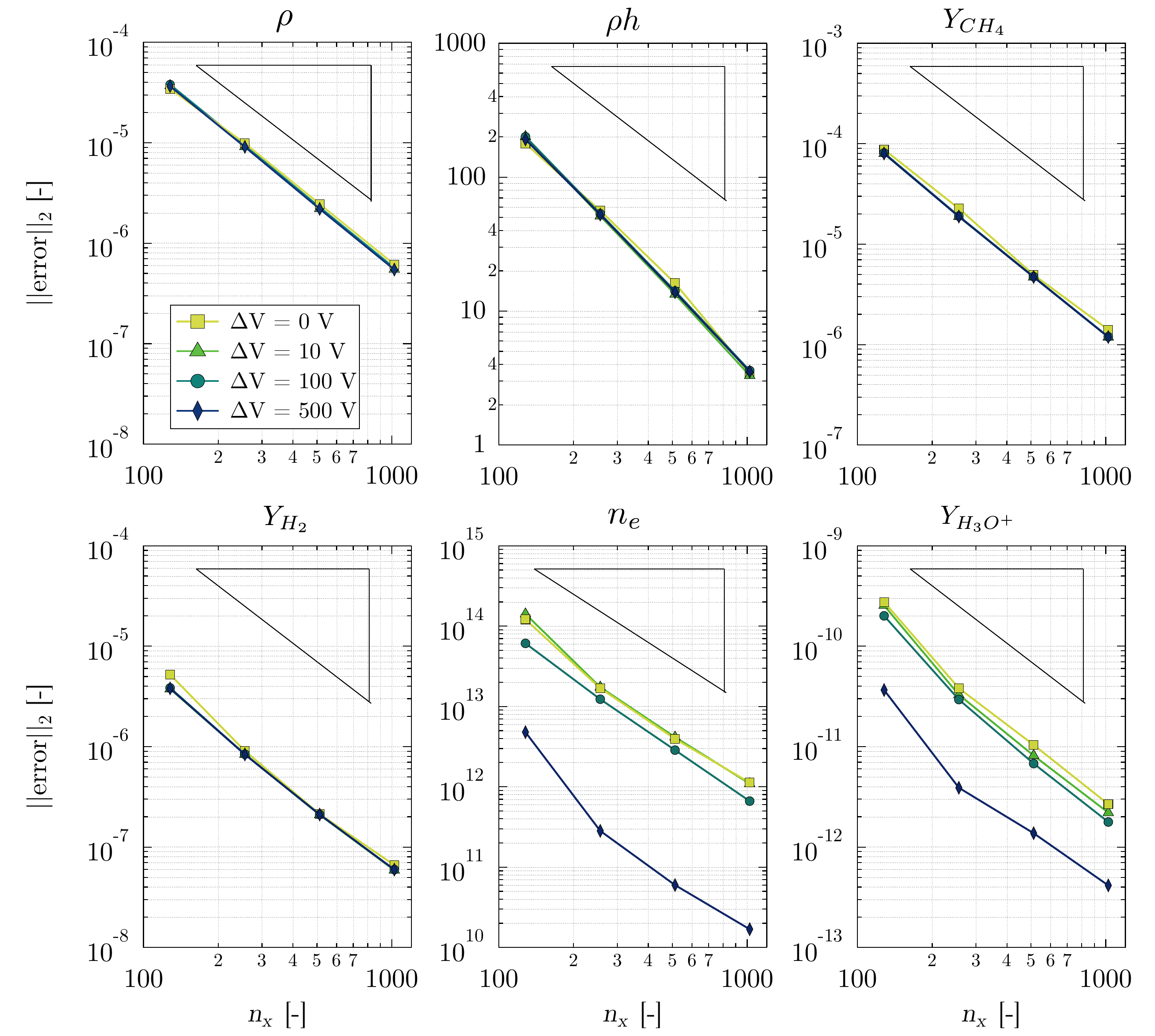}
\caption{$L^2$-norm of the error in the 1D premixed flame with four grid resolution and increasing the external electrical forcing.}
\label{fig:convergence}
\end{figure*}

In order to evaluate the convergence properties of the complete algorithm, simulations are performed halving the inlet velocity and evolving the system to a fixed time with increasing resolution, decreasing $\Delta x$ by a factor two with each refinement. The simulations are performed at a constant CFL$_{m,conv}$. The error is obtained by comparing the results at resolution $\Delta x$ with those computed with twice the resolution $\Delta x/2$. The $L^2$ norm of the error for a simulation with $n_x$ cells is:
\begin{equation}
L^2_{n_x} = \sqrt{\frac{1}{n_x}\sum_{i=1}^{n_x}\left(\varphi_i-\varphi_i^{c-f}\right)^2}
\end{equation}
where $\varphi_i^{c-f}$ is the average of the fine results onto the coarser grid. Figure~\ref{fig:convergence} shows the $L^2$ norm of the error at four grid resolutions for 6 scalars: $\rho$, $\rho h$, $Y_{CH_4}$, $Y_{H_2}$, $n_e$ and $Y_{H_3O^+}$. The slope of the error shows that second order is reached for all variables across the range of external forcing considered. The error on neutral species and mixture averaged quantities is not affected by the external forcing whereas the error on $n_e$ and $Y_{H_3O^+}$ decreases for high $\Delta V$ values (but remains a second order convergence rate). This decrease in errors indicates that the applied voltage is higher than the saturation voltage at which the charged species are drawn away from the reaction zone by the electric drift as fast as they are produced by chemical reactions.

\subsection{Steady premixed flame under DC}
\label{ssec:prem_steady}

\begin{figure*}[ht!]
\centering
\includegraphics[width=0.95\textwidth]{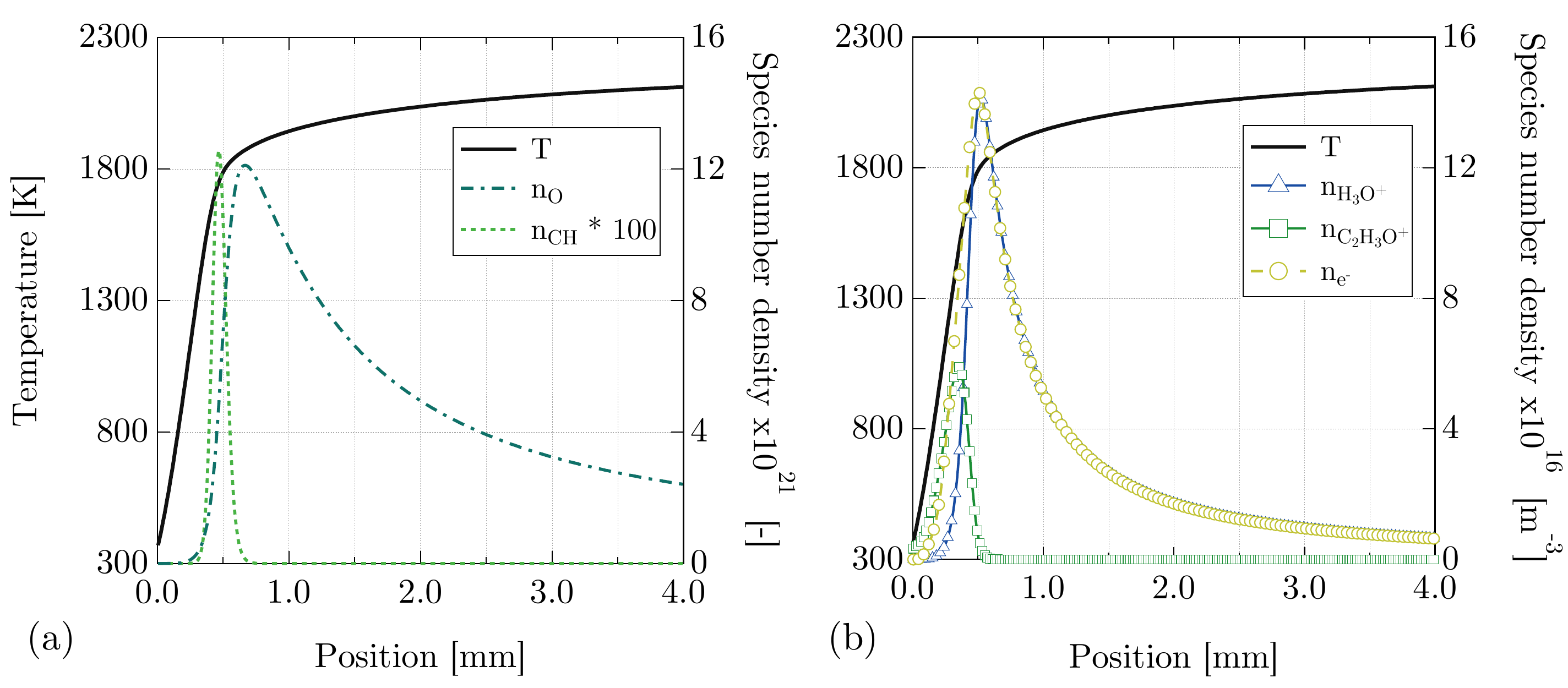}
\caption{(a) Profiles of temperature, $n_O$ and $n_{CH}$ across the flame. (b) Profiles of temperature and $n_e$, $n_{H_3O^+}$ $n_{H_3O^+}$  and $n_{C_2H_3O^+}$ across the flame. $\Delta V = 0$.}
\label{fig:flame_dV0}
\end{figure*}

Burner-stabilized, steady-state premixed methane/air flames subjected to DC electric fields have been studied using the PREMIX program in previous studies \cite{Speelman:2015,Belhi:2018}. Fig.~\ref{fig:flame_dV0} shows the temperature as well as oxygen, CH, electrons, H$_3$O$^+$ and C$_2$H$_3$O$^+$ profiles across the flame in the absence of an external electric field. Oxygen and CH are the key neutral species controlling the production rate of electrons, i.e. the number of charged particles in the flame and consequently the maximum current that can be drawn from the flame \cite{Speelman:2015, Han:2017}. Accordingly, the peak electron density in Fig.~\ref{fig:flame_dV0} is located near the corresponding maximum of CH. Note that the number density of charged species is about 5 orders of magnitude smaller than that of an intermediate radical such as O. In the absence of an external electric field, the sum of number densities of the two major cations (H$_3$O$^+$ and C$_2$H$_3$O$^+$) equals that of electrons, as ambipolar diffusion tends to balance charge separation resulting in a near electro-neutral gas.

The peak value of electron and H$_3$O$^+$ is higher than that reported in a previous study \cite{Belhi:2018} where the neutral chemical mechanism was optimized to better reproduce the CH distribution. This study showed that the GRI3.0 mechanism over-predicts the CH mass fraction, resulting in higher chemi-ionization rate and electron maximum number density. 

\begin{figure*}[ht!]
\centering
\includegraphics[width=0.7\textwidth]{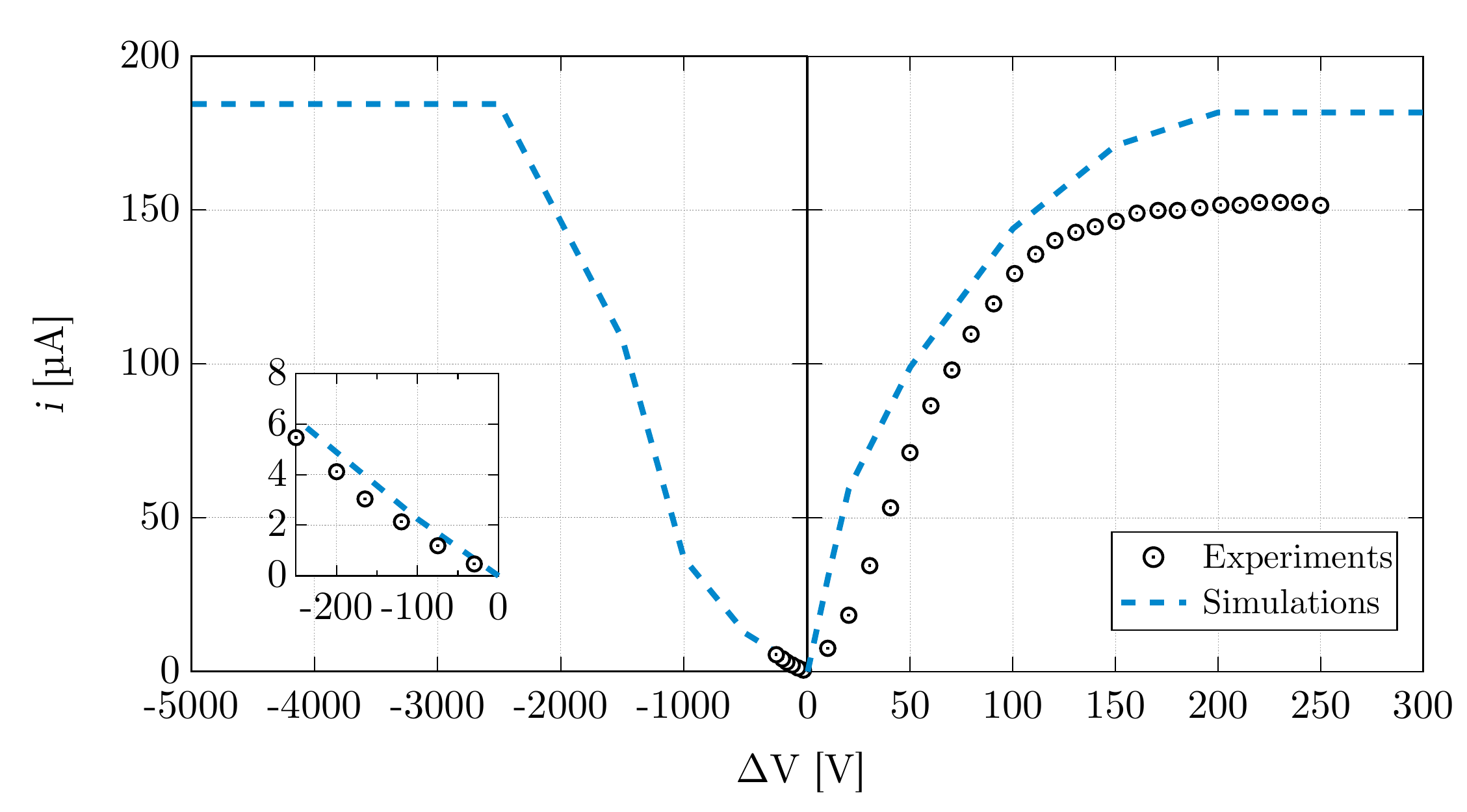}
\caption{Comparison between experimental and numerical $i$-V curves}
\label{fig:iVcurve}
\end{figure*}

Figure~\ref{fig:iVcurve} shows comparisons between experimental $i$-V curves \cite{Speelman:2015} and the present simulations. The current $i$ is evaluated by computing the charge flux carried by the charged species $m$:
\begin{equation}
\boldsymbol{J}_m = \frac{z_m}{q_e}\boldsymbol{\Gamma}_m + \frac{z_m}{q_e}\rho Y_m \boldsymbol{U}_{ef,m}
\end{equation}
and summing over positive and negative species:
\begin{align}
\boldsymbol{J}^+ & = \sum_{m \in N_p} \boldsymbol{J}_m \\
\boldsymbol{J}^- & = \sum_{m \in N_n} \boldsymbol{J}_m  + \boldsymbol{J}_e\\
\boldsymbol{J} & = \boldsymbol{J}^+ + \boldsymbol{J}^-
\end{align}
Finally, $i = \boldsymbol{J} S_{expe}$, where $S_{expe} = 7.04$ cm$^{-2}$ is the experimental cross section of the burner \cite{Speelman:2015}. Note that from the species conservation equations, in steady-state conditions $\boldsymbol{J}_m = \frac{z_m}{q_e}\dot{\omega}_m$, showing that the current drawn from the flame is directly related to the production rate of charged particles. The simulation results are consistent with the experimental data: the current increases for positive voltage until it reaches a plateau as the applied voltage exceeds a saturation value. In contrast, higher negative voltage is required to reach saturation conditions. This effect of the polarity, known as diodic effect, results from the large difference in distance between the flame and each electrode \cite{Speelman:2015}. The over-prediction of the saturation current is consistent with the fact that the mass fraction of CH is over-predicted by the GRI3.0 mechanism.

\begin{figure*}[ht!]
\centering
\includegraphics[width=0.95\textwidth]{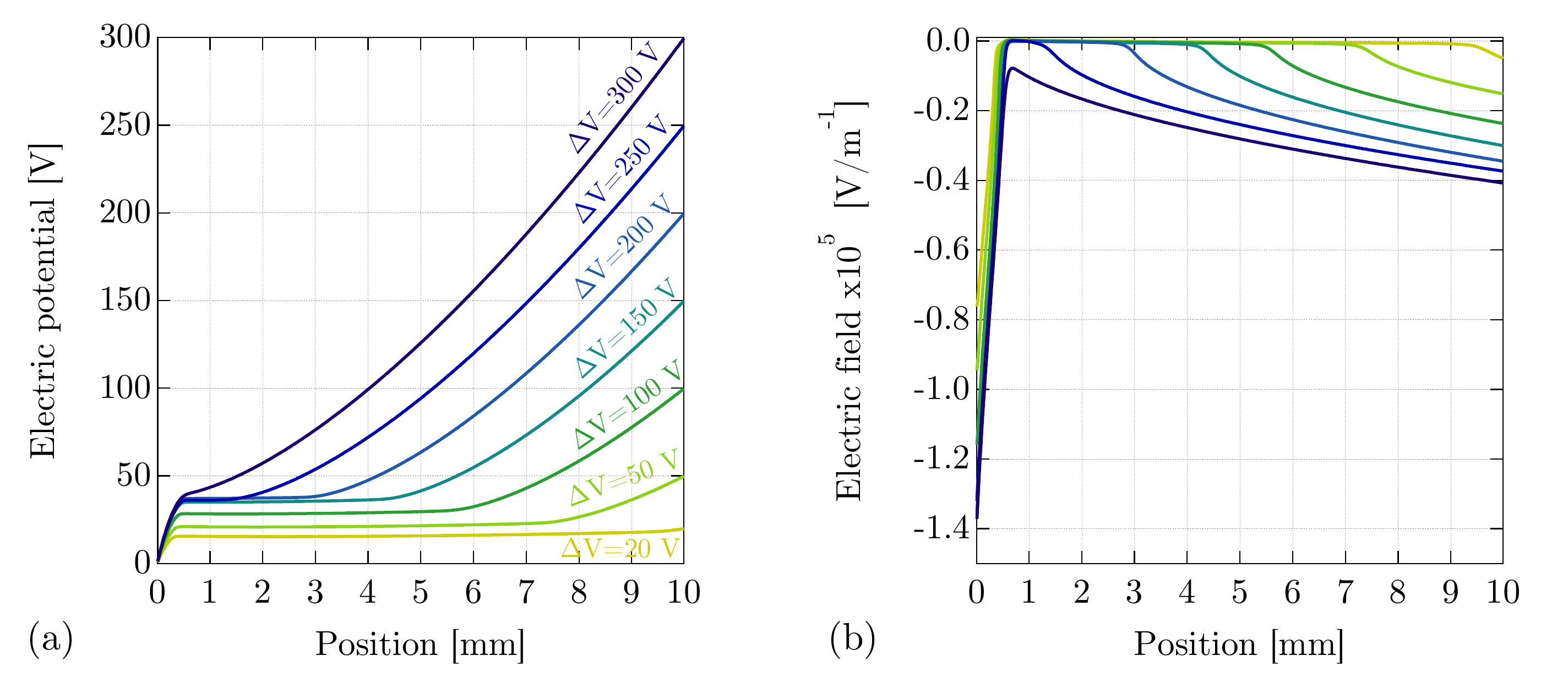}
\caption{(a) Electrostatic potential and (b) Electric field for positive values of $\Delta V$.}
\label{fig:elecfield_pos}
\end{figure*}

\begin{figure*}[ht!]
\centering
\includegraphics[width=0.95\textwidth]{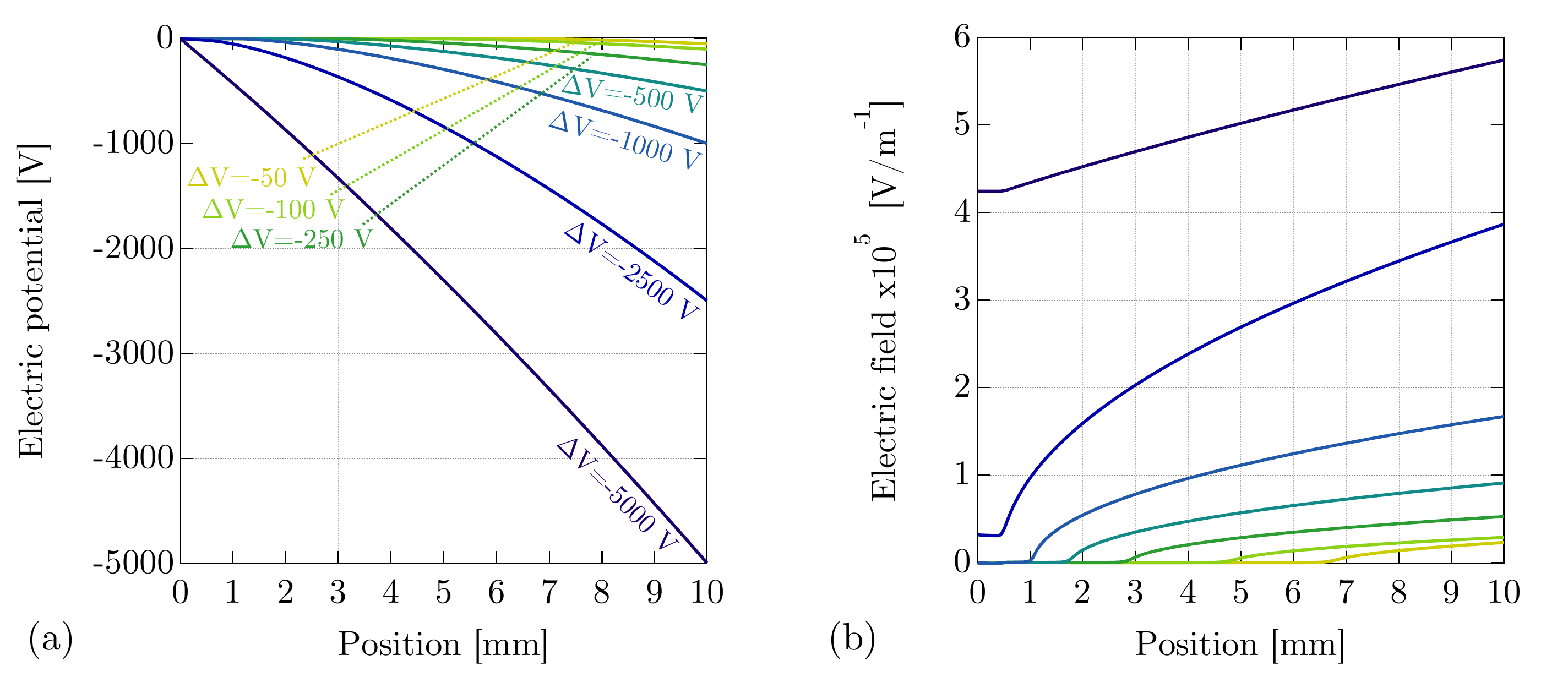}
\caption{(a) Electrostatic potential and (b) Electric field for negative values of $\Delta V$.}
\label{fig:elecfield_neg}
\end{figure*}

\begin{figure*}[ht!]
\centering
\includegraphics[width=0.95\textwidth]{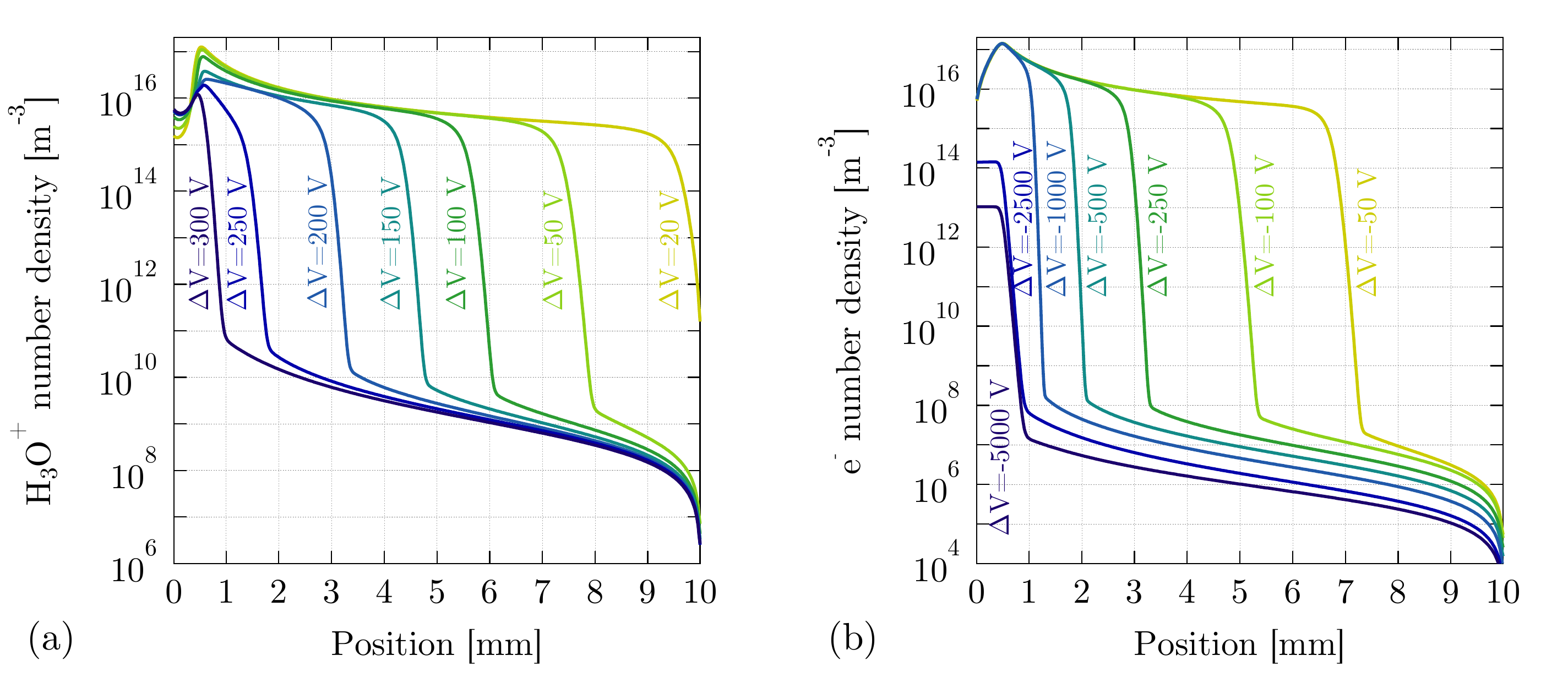}
\caption{(a) H$_3$O$^+$ profiles for positive values of $\Delta V$. (b) $e^-$ profiles for negatve values of $\Delta V$.}
\label{fig:charged_spec_sheath}
\end{figure*}

The profiles of charge particles at different values of the external voltage $\Delta V$ is presented in more detail in \Cref{fig:elecfield_pos,fig:elecfield_neg,fig:charged_spec_sheath}, which also show the steady-state profiles of electrostatic potential and electric field, for both positive and negative $\Delta V$. For sub-saturation voltages, the electric field profiles show the existence of a 'dead zone', where the electric field is close to zero and the particles are not affected by electric forces. As the external voltage intensity increases, the electrode sheath develops, eventually penetrating into the reaction zone of the flame. As saturation conditions are reached, the peak number densities of charged particles drop since they are convected away from the reaction zone as fast as they are produced through chemi-ionization. This drop is responsible for the jump of $\Delta t_{diel}$ in Fig.~\ref{fig:timestep} and the drop in $L^2$-norm of the error on charged particles in Fig.~\ref{fig:convergence}.  
 
The charged particles profiles in Fig.~\ref{fig:charged_spec_sheath} show that the proposed algorithm is able to accommodate very sharp profiles in the charged species distribution, without introducing numerical noise that would eventually lead to unstable numerical oscillations due to the strong non-linear coupling between the electrons and the electrostatic potential.

\subsection{Unsteady premixed flame}

In order to demonstrate the potential of the method to tackle unsteady simulations, the behavior of the burner-stabilized flame subjected to AC conditions is studied. Three forcing amplitudes $A_{AC}$ are considered: 100V, 1000V and 2500V respectively corresponding to conditions below both positive and negative saturation voltages, above positive and below negative saturation voltages and above both positive and negative saturation voltage. In order to estimate a characteristic relaxation time of the electrical structure of the flame, the time required for the electron and H$_3$O$^+$ to travel across the computational domain is evaluated using an averaged effective velocity $\overline{\boldsymbol{U}}_{ef}$ based on a representative mobility value for each particle and the external forcing intensity $\Delta V$. Table \ref{tab:relax_time} summarizes these relaxation times for the three forcing amplitudes considered, showing that electrons relax within a few micro-seconds, whereas it takes about a 100 times longer for the H$_3$O$^+$ to relax.

\begin{table*}[ht!]
\centering
\begin{tabular}{| c c c c | c c c c | c c c c |}
\hline
\multicolumn{12}{|c|}{Relaxation time [s]}\\\hline
\multicolumn{4}{|c|}{$\Delta V$ = 100V} & \multicolumn{4}{c|}{$\Delta V$ = 1000V} & \multicolumn{4}{c|} {$\Delta V$ = 2500V}\\\hline
\multicolumn{2}{|c}{$\tau_{e^-}$} & \multicolumn{2}{c|}{$\tau_{H_3O^+}$} & \multicolumn{2}{c}{$\tau_{e^-}$} & \multicolumn{2}{c|}{$\tau_{H_3O^+}$} & \multicolumn{2}{c}{$\tau_{e^-}$} & \multicolumn{2}{c|}{$\tau_{H_3O^+}$} \\
\multicolumn{2}{|c}{2.5e$^{-6}$} & \multicolumn{2}{c|}{2.5e$^{-4}$} & \multicolumn{2}{c}{2.5e$^{-7}$} & \multicolumn{2}{c|}{2.5e$^{-5}$} & \multicolumn{2}{c}{1.0e$^{-7}$} & \multicolumn{2}{c|}{1.0e$^{-5}$} \\\hline
\end{tabular}
\caption{Estimated relaxation time of electrons and H$_3$O$^+$ for increasing external forcing intensity $\Delta V$.}\label{tab:relax_time}
\end{table*}

These relaxation time scales can be compared to the half period of an AC forcing $\tau_{AC}/2$ to distinguish several regimes (for a fixed value of $\Delta V$): 1) for low forcing frequency, $f_{AC}$, both electrons and ions remain in quasi-equilibrium with the instantaneous potential difference and the charged particles profiles are close to the corresponding steady-states; 2) for higher $f_{AC}$, the electrons are close to equilibrium, but the slower ions do not reach steady-state, changing the current drawn from the flame and possibly inducing an asymmetric ionic wind due to the diodic effect; and 3) for very high $f_{AC}$ the ions are too slow to respond to the change in external electric potential and the ionic wind (mainly due to the motion of ions) becomes negligible. In practice, only the first two regimes are of interest to study the effect of ionic wind on the flame behavior. Additionally, in the first regime, the flame structures are expected to remain close to the ones described in Section \ref{ssec:prem_steady} so that we will focus on the second regime by considering $f_{AC}$ listed in Table \ref{tab:forcing_freq}.


\begin{table*}[ht!]
\centering
\begin{tabular}{| l | c c c c  |}
\hline
Case & $f_{AC}$ [Hz] & $\tau_{AC}/2$ [s] & $\tau_{AC}/2 \tau_{e^-}$ & $\tau_{AC}/2\tau_{H_3O^+}$ \\\hline
AC1k    & 1000   & 5e$^{-4}$ & 2000 & 20 \\
AC2.5k & 2500   & 2e$^{-4}$ & 800 & 8 \\
AC5k    & 5000   & 1e$^{-4}$ & 400 & 4 \\
AC10k  & 10000 & 5e$^{-5}$ & 200 & 2 \\
AC25k  & 25000 & 2e$^{-5}$ & 80   & 0.8 \\
AC50k  & 50000 & 1e$^{-5}$ & 40   & 0.4 \\
AC100k  & 100000 & 5e$^{-6}$ & 20   & 0.2 \\\hline
\end{tabular}
\caption{Considered forcing frequencies and half periods for the AC cases along with ratios to the characteristic relaxation time of electrons and H$_3$O$^+$ at $A_{AC} = 1000$ V.}
\label{tab:forcing_freq}
\end{table*}

\begin{figure*}[ht!]
\centering
\includegraphics[width=0.92\textwidth]{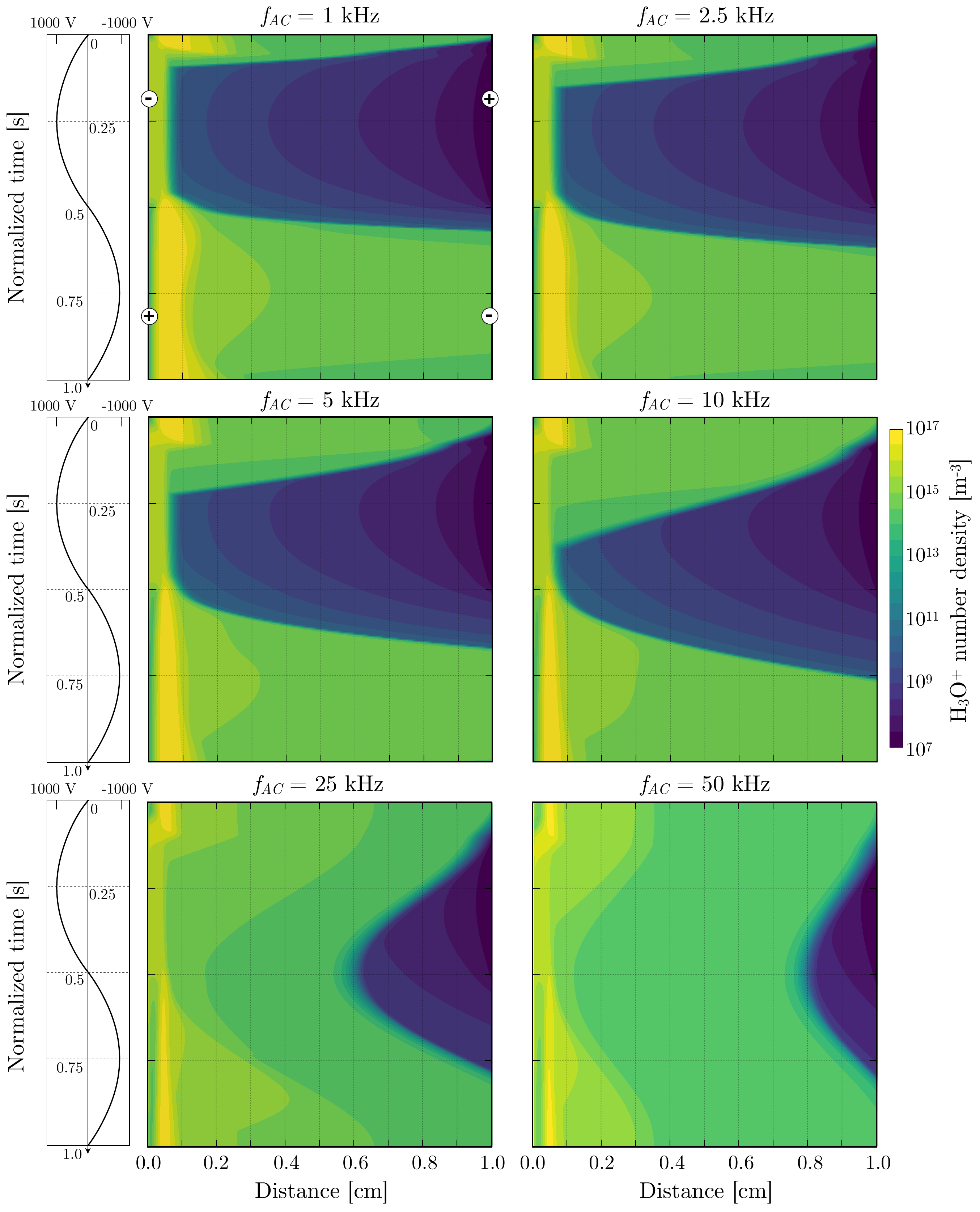}
\caption{Space (horizontal) and time (vertical) evolution of the H$_3$O$^+$ number density at six AC forcing frequency and $A_{AC} = 1000$ V. Time is normalized by the forcing period $\tau_{AC}$.}
\label{fig:transient_H3Op}
\end{figure*}

The temporal evolution of the H$_3$O$^+$ profile during a statistically steady period of the AC forcing is shown in Fig.~\ref{fig:transient_H3Op} for the seven values of $f_{AC}$ considered at a constant forcing amplitude of 1000 V. Each plot shows the evolution of the one-dimensional H$_3$O$^+$ number density profile (horizontal direction) as function of the normalized time $t^{*} = t / \tau_{AC}$ (vertical direction, from top to bottom). A few periods are necessary before reaching statistically steady oscillations. Note that, these plots confirm that the proposed algorithm is able to smoothly capture the fast motion of steep charged species fronts. 

Figure~\ref{fig:transient_H3Op} shows that for the initially positive polarity, the development of the cathode sheath is qualitatively similar to the steady states depicted in Fig.~\ref{fig:charged_spec_sheath}(a). For low forcing frequency, positive saturation conditions are reached for most of the cycle first half-period. As the forcing frequency increases, the cathode sheath is no longer able to fully develop and H$_3$O$^+$ depletion near the right boundary of the computational domain remains during part the second half of the cycle, even though the polarity is reversed. Additionally, the peak value of H$_3$O$^+$ is found to decrease with increasing frequency, indicating that the charged particle profiles are no longer able to relax to the forcing free profiles while the instantaneous voltage is close to zero.

\begin{figure*}[ht!]
\centering
\includegraphics[width=0.95\textwidth]{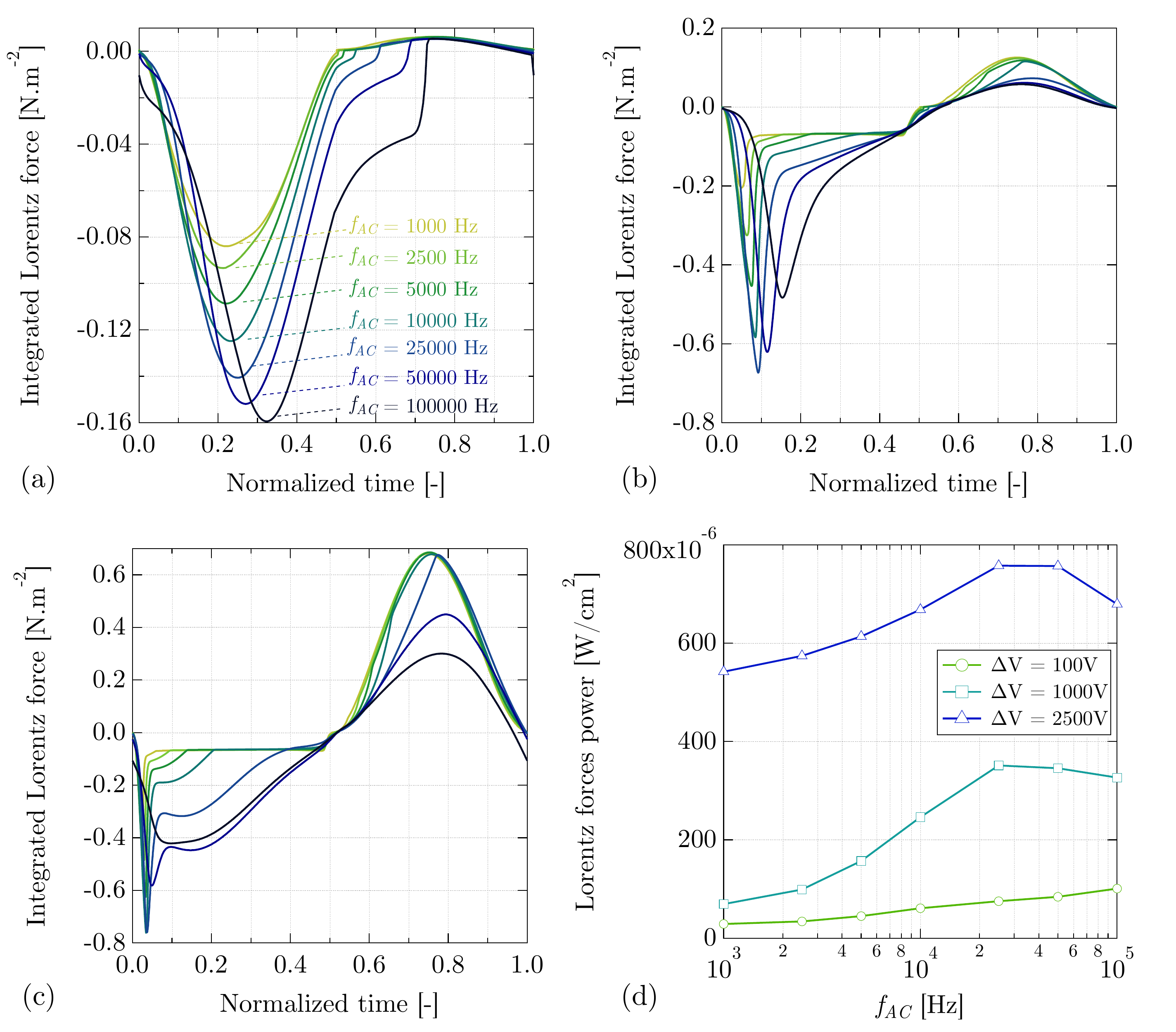}
\caption{(a)-(c): temporal evolution of the integrated Lorentz forces $F_{Lorentz}$. Time is normalized by the AC period. (d) : evolution of the Lorentz forces average work as function of $f_{AC}$.}
\label{fig:temporals}
\end{figure*}

To analyze the effect of the forcing frequency and amplitude on the ionic wind effect, the integral of the Lorentz forces appearing the momentum equation (\ref{eq:mom}) across the computational domain is computed:
\begin{equation}
F_{Lorentz} = \int_x \rho \sum_{m+e} z_m Y_{m} \boldsymbol{E} \; \text{d}x
\end{equation}
corresponding to the force per unit area. Additionally, to evaluate the effect of this force on the flame, the average work of the Lorentz forces $W_{Lorentz}$ over a period is evaluated from the Ohmic heating term appearing in the energy equation (\ref{eq:nrg}). Figures~\ref{fig:temporals}(a-c) show the temporal evolution of $F_{Lorentz}$ during one AC forcing period for the three values of the forcing amplitude while Fig.~\ref{fig:temporals}(d) shows the evolution of $W_{Lorentz}$ as function of the frequency for different values of $A_{AC}$. For small forcing amplitude (below both positive and negative saturation values) the proximity of the flame to the anode, also responsible for the diodic effect, results in an overall negative Lorentz force, the work of which increases with increasing frequency. As the forcing amplitude increases, the positive Lorentz force during the second half of forcing cycle becomes more important. In these conditions, increasing the forcing frequency results in an increase of the averaged work generated by the Lorentz force, up to a critical frequency above which the electric field is no longer able to penetrate into the flame and the work begin to decrease. These results indicate that, as in the DC cases, the effect of the AC electric field not only depends upon the forcing frequency and amplitude, but also the flame position compared to the electrodes.

\section{Conclusion}

This work proposes a new numerical strategy to include the motion of charged particles in simulations of low Mach number reactive flows in the presence of electric fields. We have found that to overcome the stringent time-step constraint imposed by fast electrons and their coupling with the electrostatic potential equation, a non-linear implicit solution of the system of equations governing these two quantities is necessary. Keeping in mind the need for an efficient methodology in large scale (multi-dimensional) computations, we have developed an algorithm that introduces a JFNK solver within the SDC iterations developed for classical reactive flow simulation. To obtain good performance, we constructed a preconditioner based on the Schur decomposition of the Jacobian matrix for the electrons/electrostatic potential system. An approximation of the Schur complement of the Jacobian matrix is proposed enabling use of multi-grid method to approximate the inverse of the preconditioner in the iterative linear solve. 

We demonstrated on one-dimensional burner-stabilized premixed flame configurations, that second-order accuracy is reached for all the transported variables and for a wide range of external electric forcing. The numerical results compare well with experimental data regarding the current-voltage characterization of the flame (given the uncertainty on the chemical mechanism) and detailed analysis of the charged particles profiles are consistent with previous studies using steady-state one-dimensional solvers.

The proposed strategy is currently being implemented in the low Mach number reactive flow solver PeleLM, which is based on the block-structured adaptive mesh refinement library AMReX. The resulting unique numerical tool will allow us to investigate realistic engineering applications of electric field controlled flames that have so far not been possible.

\section*{Acknowledgements}
This work was supported by the U.S. Department of Energy, Office of Science, Office of Advanced Scientific Computing Research, Applied Mathematics Program (under Award Number DE-SC0008271 and under contract No. DE-AC02-05CH11231).

\clearpage

\bibliography{scibib}

\begin{thebibliography}{10}

\bibitem{Fialkov:1997}
A.~B. Fialkov.
\newblock Investigations on ions in flames.
\newblock {\em Progress in Energy and Combustion Science}, 23(5):399 -- 528,
  1997.

\bibitem{Starikovskiy:2013}
A.~Starikovskiy and N.~Aleksandrov.
\newblock Plasma-assisted ignition and combustion.
\newblock {\em Progress in Energy and Combustion Science}, 39(1):61 -- 110,
  2013.

\bibitem{Adams:1995}
J.~T. Adams, J.~E. Bohan~Jr, and R.~W. Simons.
\newblock Flame rectification sensor employing pulsed excitation, 1995.
\newblock US Patent 5,472,336.

\bibitem{Calcote:1957}
H.F. Calcote.
\newblock Mechanisms for the formation of ions in flames.
\newblock {\em Combustion and Flame}, 1(4):385 -- 403, 1957.

\bibitem{Goodings:1979a}
J.M. Goodings, D.K. Bohme, and Chun-Wai Ng.
\newblock Detailed ion chemistry in methane-oxygen flames. {I}. positive ions.
\newblock {\em Combustion and Flame}, 36:27 -- 43, 1979.

\bibitem{Goodings:1979b}
J.M. Goodings, D.K. Bohme, and Chun-Wai Ng.
\newblock Detailed ion chemistry in methane-oxygen flames. {II}. negative ions.
\newblock {\em Combustion and Flame}, 36:45 -- 62, 1979.

\bibitem{Jaggers:1971}
HC~Jaggers and A~Von~Engel.
\newblock The effect of electric fields on the burning velocity of various
  flames.
\newblock {\em Combustion and Flame}, 16(3):275--285, 1971.

\bibitem{Tewari:1975}
GP~Tewari and JR~Wilson.
\newblock An experimental study of the effects of high frequency electric
  fields on laser-induced flame propagation.
\newblock {\em Combustion and Flame}, 24:159--167, 1975.

\bibitem{Marcum:2005}
S.D. Marcum and B.N. Ganguly.
\newblock Electric-field-induced flame speed modification.
\newblock {\em Combustion and Flame}, 143(1):27 -- 36, 2005.

\bibitem{Won:2008}
S.H. Won, S.K. Ryu, M.K. Kim, M.S. Cha, and S.H. Chung.
\newblock Effect of electric fields on the propagation speed of tribrachial
  flames in coflow jets.
\newblock {\em Combustion and Flame}, 152(4):496 -- 506, 2008.

\bibitem{Kim:2011}
M.K. Kim, S.H. Chung, and H.H. Kim.
\newblock Effect of {AC} electric fields on the stabilization of premixed
  bunsen flames.
\newblock {\em Proceedings of the Combustion Institute}, 33(1):1137 -- 1144,
  2011.

\bibitem{Cessou:2012}
A.~Cessou, E.~Varea, K.~Criner, G.~Godard, and P.~Vervisch.
\newblock Simultaneous measurements of {OH}, mixture fraction and velocity
  fields to investigate flame stabilization enhancement by electric field.
\newblock {\em Experiments in Fluids}, 52(4):905--917, 2012.

\bibitem{Kono:1989}
M~Kono, FB~Carleton, AR~Jones, and FJ~Weinberg.
\newblock The effect of nonsteady electric fields on sooting flames.
\newblock {\em Combustion and flame}, 78(3-4):357--364, 1989.

\bibitem{Vatazhin:1995}
A.~B. Vatazhin, V.~A. Likhter, V.~A. Sepp, and V.~I. Shul'gin.
\newblock Effect of an electric field on the nitrogen oxide emission and
  structure of a laminar propane diffusion flame.
\newblock {\em Fluid Dynamics}, 30(2):166--174, April 1995.

\bibitem{Saito:1999}
M.~Saito, T.~Arai, and M.~Arai.
\newblock Control of soot emitted from acetylene diffusion flames by applying
  an electric field.
\newblock {\em Combustion and Flame}, 119(3):356 -- 366, 1999.

\bibitem{Jones:1972}
F.~L Jones, P.~M Becker, and R.~J Heinsohn.
\newblock A mathematical model of the opposed-jet diffusion flame: Effect of an
  electric field on concentration and temperature profiles.
\newblock {\em Combustion and flame}, 19(3):351--362, 1972.

\bibitem{Pedersen:1993}
T.~Pedersen and R.~Brown.
\newblock Simulation of electric field effects in premixed methane flames.
\newblock {\em Combustion and Flame}, 94(4):433--448, 1993.

\bibitem{Prager:2007}
J.~Prager, U.~Riedel, and J.~Warnatz.
\newblock Modeling ion chemistry and charged species diffusion in lean
  methane-oxygen flames.
\newblock {\em Proceedings of the Combustion Institute}, 31(1):1129--1137,
  2007.

\bibitem{Peerlings:2013}
L.~Peerlings, V.~Kornilov, and P.~de~Goey.
\newblock Flame ion generation rate as a measure of the flame thermo-acoustic
  response.
\newblock {\em Combustion and Flame}, 160(11):2490--2496, 2013.

\bibitem{Speelman:2015}
N.~Speelman, M.~Kiefer, D.~Markus, U.~Maas, L.~de~Goey, and J.A. van Oijen.
\newblock Validation of a novel numerical model for the electric currents in
  burner-stabilized methane--air flames.
\newblock {\em Proceedings of the Combustion Institute}, 35(1):847--854, 2015.

\bibitem{Belhi:2017}
M.~Belhi, B.~J. Lee, F.~Bisetti, and H.~G. Im.
\newblock A computational study of the effects of {DC} electric fields on
  non-premixed counterflow methane-air flames.
\newblock {\em Journal of Physics D: Applied Physics}, 50(49):494005, 2017.

\bibitem{Han:2017}
J.~Han, M.~Belhi, T.~Casey, F.~Bisetti, H.~G. Im, and J.-Y. Chen.
\newblock The i-{V} curve characteristics of burner-stabilized premixed flames:
  detailed and reduced models.
\newblock {\em Proceedings of the Combustion Institute}, 36(1):1241--1250,
  2017.

\bibitem{Bisetti:2012}
F.~Bisetti and M.~El~Morsli.
\newblock Calculation and analysis of the mobility and diffusion coefficient of
  thermal electrons in methane/air premixed flames.
\newblock {\em Combustion and flame}, 159(12):3518--3521, 2012.

\bibitem{Han:2015}
J.~Han, M.~Belhi, F.~Bisetti, and S.~Mani~Sarathy.
\newblock Numerical modelling of ion transport in flames.
\newblock {\em Combustion Theory and Modelling}, 19(6):744--772, 2015.

\bibitem{Papac:2008}
M.J. Papac and D.~Dunn-Rankin.
\newblock Modelling electric field driven convection in small combustion
  plasmas and surrounding gases.
\newblock {\em Combustion Theory and Modelling}, 12(1):23--44, 2008.

\bibitem{Renzo:2018}
M.~Di Renzo, J.~Urzay, P.~De Palma, M.~D. de~Tullio, and G.~Pascazio.
\newblock The effects of incident electric fields on counterflow diffusion
  flames.
\newblock {\em Combustion and Flame}, 193:177 -- 191, 2018.

\bibitem{Yamashita:2009}
K.~Yamashita, S.~Karnani, and D.~Dunn-Rankin.
\newblock Numerical prediction of ion current from a small methane jet flame.
\newblock {\em Combustion and Flame}, 156(6):1227--1233, 2009.

\bibitem{Belhi:2010}
M.~Belhi, P.~Domingo, and P.~Vervisch.
\newblock Direct numerical simulation of the effect of an electric field on
  flame stability.
\newblock {\em Combustion and flame}, 157(12):2286--2297, 2010.

\bibitem{Belhi:2013}
M.~Belhi, P.~Domingo, and P.~Vervisch.
\newblock Modelling of the effect of {DC} and {AC} electric fields on the
  stability of a lifted diffusion methane/air flame.
\newblock {\em Combustion Theory and Modelling}, 17(4):749--787, 2013.

\bibitem{Belhi:2019}
M. Belhi, B.~J. Lee, M.~S. Cha, and H.~G Im.
\newblock Three-dimensional simulation of ionic wind in a laminar premixed
  bunsen flame subjected to a transverse dc electric field.
\newblock {\em Combustion and Flame}, 202:90--106, 2019.

\bibitem{Hagelaar:2000}
G.~Hagelaar and G.~Kroesen.
\newblock Speeding up fluid models for gas discharges by implicit treatment of
  the electron energy source term.
\newblock {\em Journal of Computational Physics}, 159(1):1--12, 2000.

\bibitem{Nonaka:2012}
A.~Nonaka, J.B. Bell, M.S. Day, C.~Gilet, A.S. Almgren, and M.L. Minion.
\newblock A deferred correction coupling strategy for low {M}ach number flow
  with complex chemistry.
\newblock {\em Combustion Theory and Modeling}, 16(6):1053--1088, 2012.

\bibitem{Day:2000}
M.S. Day and J.B. Bell.
\newblock Numerical simulation of laminar reacting flows with complex
  chemistry.
\newblock {\em Combustion Theory Modelling}, 4:535--556, 2000.

\bibitem{Majda:1985}
A. Majda and J. Sethian.
\newblock The derivation and numerical solution of the equations for zero mach
  number combustion.
\newblock {\em Combustion science and technology}, 42(3-4):185--205, 1985.

\bibitem{ref19}
A.~Ern and V.~Giovangigli.
\newblock {EGLIB}: a general-purpose fortran library for multicomponent
  transport property evaluation.
\newblock {\em CERMICS Internal Report}, 96-51, 1996.

\bibitem{ref20}
Y.~Roichman and N.~Tessler.
\newblock Generalized {E}instein relation for disordered
  semiconductors---implications for device performance.
\newblock {\em Applied Physics Letters}, 80(11), 2002.

\bibitem{SmithGRI:2000}
GP~Smith, DM~Golden, M~Frenklach, NW~Moriarty, B~Eiteneer, M~Goldenberg,
  CT~Bowman, RK~Hanson, S~Song, WC~Gardiner~Jr, et~al.
\newblock Gri-mech 3.0, 2000.
\newblock {\em URL http://www. me. berkeley. edu/gri\_mech}.

\bibitem{Belhi:2018}
M. Belhi, J. Han, T. A. Casey, J-Y. Chen, H.~G. Im, S.~Mani
  Sarathy, and F. Bisetti.
\newblock Analysis of the current--voltage curves and saturation currents in
  burner-stabilised premixed flames with detailed ion chemistry and transport
  models.
\newblock {\em Combustion Theory and Modelling}, 22(5):939--972, 2018.

\bibitem{Burcat:2006}
A.~Burcat.
\newblock Burcat's thermodynamic data: Ideal gas thermodynamic data in
  polynomial form for combustion and air pollution use.

\bibitem{Mason:1988}
E.~A Mason and E.~W McDaniel.
\newblock {\em Transport properties of ions in gases}.
\newblock Wiley-Interscience, New-York, 1988.

\bibitem{Bisetti:2014}
F. Bisetti and M. El~Morsli.
\newblock Kinetic parameters, collision rates, energy exchanges and transport
  coefficients of non-thermal electrons in premixed flames at sub-breakdown
  electric field strengths.
\newblock {\em Combustion Theory and Modelling}, 18(1):148--184, 2014.

\bibitem{Hagelaar:2005}
GJM Hagelaar and LC~Pitchford.
\newblock Solving the boltzmann equation to obtain electron transport
  coefficients and rate coefficients for fluid models.
\newblock {\em Plasma Sources Science and Technology}, 14(4):722, 2005.

\bibitem{Nonaka:2018}
A. Nonaka, M.~S Day, and J.~B Bell.
\newblock A conservative, thermodynamically consistent numerical approach for
  low mach number combustion. part i: Single-level integration.
\newblock {\em Combustion Theory and Modelling}, 22(1):156--184, 2018.

\bibitem{Dutt:2000}
A.~Dutt, L.~Greengard, and V.~Rokhlin.
\newblock Spectral deferred correction methods for ordinary differential
  equations.
\newblock {\em BIT Numerical Mathematics}, 40(2):241--266, 2000.

\bibitem{Bourlioux:2003}
A. Bourlioux, A.~T Layton, and M.~L Minion.
\newblock High-order multi-implicit spectral deferred correction methods for
  problems of reactive flow.
\newblock {\em Journal of Computational Physics}, 189(2):651--675, 2003.

\bibitem{Layton:2004}
A.~T Layton and M.~L Minion.
\newblock Conservative multi-implicit spectral deferred correction methods for
  reacting gas dynamics.
\newblock {\em Journal of Computational Physics}, 194(2):697--715, 2004.

\bibitem{Knoll:2004}
D.~A Knoll and D.~E Keyes.
\newblock Jacobian-free newton--krylov methods: a survey of approaches and
  applications.
\newblock {\em Journal of Computational Physics}, 193(2):357--397, 2004.

\bibitem{Dennis:1996}
J.~E Dennis~Jr and R.~B Schnabel.
\newblock {\em Numerical methods for unconstrained optimization and nonlinear
  equations}, volume~16.
\newblock Siam, 1996.

\bibitem{Saad:1986}
Y. Saad and M.~H Schultz.
\newblock Gmres: A generalized minimal residual algorithm for solving
  nonsymmetric linear systems.
\newblock {\em SIAM Journal on scientific and statistical computing},
  7(3):856--869, 1986.

\bibitem{Heroux:2005}
M.~A Heroux, R.~A Bartlett, V.~E Howle, Robert~J Hoekstra,
  Jonathan~J Hu, Tamara~G Kolda, Richard~B Lehoucq, Kevin~R Long, Roger~P
  Pawlowski, Eric~T Phipps, et~al.
\newblock An overview of the trilinos project.
\newblock {\em ACM Transactions on Mathematical Software (TOMS)},
  31(3):397--423, 2005.

\bibitem{Cantera:2017}
D.~G. Goodwin, R.~L. Speth, H.~K. Moffat, and B.~W. Weber.
\newblock Cantera: An object-oriented software toolkit for chemical kinetics,
  thermodynamics, and transport processes.
\newblock \url{https://www.cantera.org}, 2017.
\newblock Version 2.3.0.

\bibitem{Briggs:2000}
W.~L Briggs, S.~F McCormick, et~al.
\newblock {\em A multigrid tutorial}, volume~72.
\newblock Siam, 2000.

\end{thebibliography}
\bibliographystyle{tfq}

\clearpage

\end{document}